\documentclass[%
 pre,
superscriptaddress,
twocolumn,
 amsmath,amssymb,
 aps,
10pt]{revtex4-2}

\usepackage{graphicx}
\usepackage{dcolumn}
\usepackage{bm}
\usepackage{hyperref}
\usepackage{comment}
\usepackage{xcolor}

\begin{document}

\title{Pulling apart the mechanisms that lead to jammed knitted fabrics}

\author{Sarah E.~Gonzalez}
\affiliation{School of Physics, Georgia Institute of Technology, Atlanta, Georgia 30332, USA}
\author{Michael S.~Dimitriyev}
\affiliation{Department of Materials Science and Engineering, Texas A\&M University, College Station, Texas 77843, USA}
\author{A. Patrick Cachine}
\affiliation{School of Physics, Georgia Institute of Technology, Atlanta, Georgia 30332, USA}
\author{Elisabetta A.~Matsumoto}\email{sabetta@gatech.edu}
\affiliation{School of Physics, Georgia Institute of Technology, Atlanta, Georgia 30332, USA}
\affiliation{International Institute for Sustainability with Knotted Chiral Meta Matter (WPI-SKCM2), Hiroshima University, Higashihiroshima 739-8526, Japan}

\date{\today}

\begin{abstract}

We investigate the mechanical behavior of \emph{jammed} knitted fabrics, where geometric confinement leads to an initially stiff mechanical response that softens into low stiffness behavior with additional applied stress.
We show that the jammed regime is distinguished by changes in yarn geometry and contact rearrangement that must occur to allow the individual stitches to stretch.
These rearrangements allow for the relaxation of high residual stresses that are present within jammed fabric, altering the low-strain response.
We demonstrate that fabric jamming can be induced or reduced by changing either the constituent yarn or the fabric manufacturing parameters. 
Analysis of experimental samples shows that changing the ``stitch size" in manufacturing affects both the \textit{in situ} yarn radius and the length of yarn per stitch, both of which affect jamming. 

\end{abstract}

\maketitle

\section{Introduction}

Granular systems are ubiquitous in nature and manufacturing, from sandy beaches to corn-filled silos, and they exhibit some common behaviors.
Generally, jamming is the process through which an amorphous or disordered system attains a yield stress through changing properties such as density or temperature \cite{liu2001jamming, ohern2002}.
This disorder differentiates jamming from crystallization, which gives rise to an ordered, equilibrium state.
For athermal hard spheres, the onset of jamming signifies the sudden transition from zero to non-zero pressure response when the granular system is compressed \cite{ohern2002}.
A universal feature of granular systems is a \emph{jamming transition}, where a system which has fluid-like behavior -- flowing under constant stress -- develops compressional rigidity and behaves like a solid under constant stress.

The transition to, or from, a jammed state often occurs due to changes in constraints on the constituent particles.
For hard particles, these constraints arise from contacts between neighbors, and the jamming transition can be rationalized by simple constraint counting and thus characterized by the average number of contacts per particle \cite{song_phase_2008}.
By contrast, the equilibrium configurations of ``elasto-granular'' systems that combine hard particle jamming with fiber elasticity involve non-local, compressive \textit{and} tensile stress distributions, giving rise to new mechanical behaviors \cite{guerra_emergence_2021, dreier_beaded_2024}.
For moderately deformable systems, like foams, the number of contacts still plays a large role in the characterization of the jamming transition \cite{brujic2007, Katgert_2010}.
In highly deformable systems -- where the deformation is of the order of the particle size -- additional constraints, such as particle shape \cite{park_unjamming_2015, boromand2018}, contribute to the onset of jammed behavior.

Jamming has been introduced as a particularly useful property in mechanical metamaterials and soft robotics, where the variable stiffness of the material leads to novel applications \cite{weiner_mechanics_2020, aktas2020, kang_soft_2023} such as soft grippers that conform around an object and utilize jamming to rigidify and move the object \cite{brown2010}.
In this vein, we examine jamming in a non-granular context,
where the continuous, flexible yarn that is initially confined in tightly-knitted fabrics must rearrange in order to support higher stresses.

\begin{figure*}[t]
    \centering
    \includegraphics[width = 17.2cm]{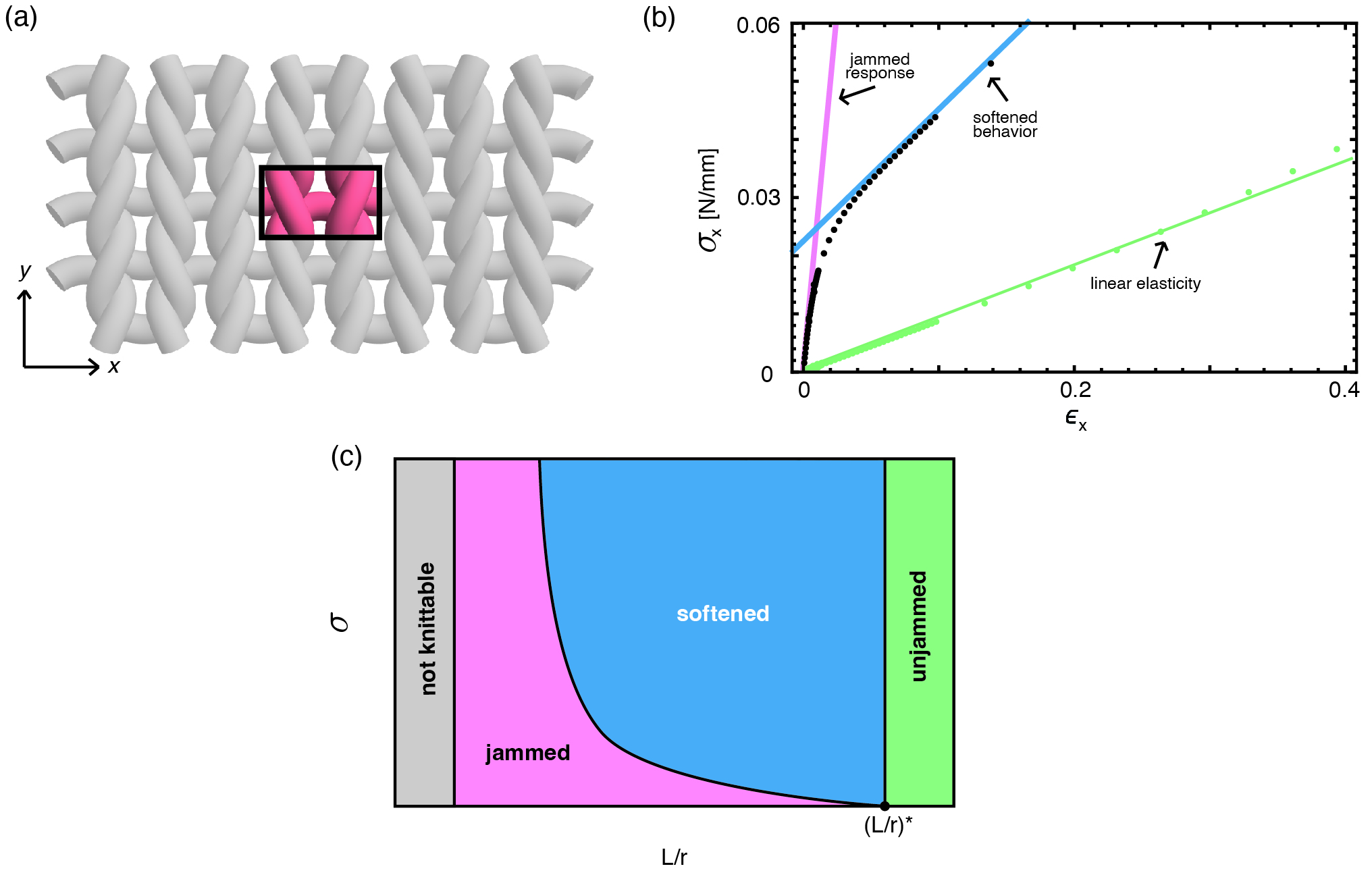}
    \caption{(a) Simulated knitted fabric with the unit cell highlighted.
    This graphic is made from a simulation where $k=0.1~{\rm mN/mm^{-2}}$,  $r=0.74~{\rm mm}$, and $L=13.0~{\rm mm}$.
    However, we rendered the image with a yarn radius $r=0.5~{\rm mm}$ instead of the simulation value to better display the yarn structure.
    (b) Stress versus strain in the x-direction for two different simulated fabrics: a jammed sample where $L=10.5~{\rm mm}$ in black and an unjammed sample where $L=12.2~{\rm mm}$ in green.
    The jammed regime is seen as a region of high tangent modulus at low strain, as indicated by the pink line.
    The softened regime is indicated with the blue line.
    (c) Schematic of jammed knitted fabric phase diagram.
    For some possible combinations of yarn length $L$ and yarn radius $r$, the  fabric is not knittable, indicated by the grey region.
    When the fabric is knittable and $L/r$ is smaller than some critical threshold $(L/r)^*$, there are some applied stresses that show jammed responses and some that show softened elastic behavior.
    When $L/r > (L/r)^*$, the fabric does not exhibit jammed behavior even for small stresses.
    We are unable to quantify $(L/r)^*$ using our simulation methods, but it is discussed in more detail in Supplementary Section \textcolor{black}{VIII}.}
    \label{fig:f1}
\end{figure*}

Knitting is a fabric manufacturing method that creates some of the most commonly used clothing items, including t-shirts and socks.
Knitted fabrics are comprised of a network of interlocked loops formed by a continuous piece of yarn (Fig.~\ref{fig:f1}a).
In weft knitting, a single strand of yarn forms a series of loops horizontally within a row.
In the next row, a new series of loops are threaded through the loops on the prior row.
Each loop of yarn is called a stitch. Combinations of stitches that have different topologies creates to different elastic responses and textures in the final fabric \cite{leaf_interactive_2018,singal_programming_2024}.
Fabrics made entirely of knit stitches are called \emph{stockinette fabric}, or \emph{jersey}, and is the fabric type found in the body of t-shirts.
This is the type of fabric we will study here, and it is the stiffest of the four most common knitted fabric types \cite{singal_programming_2024, ding_unravelling_2024}.

The term ``jamming,'' as applied to knitted textiles, was coined by Postle~\cite{Postle2002} to describe an anomalous stiff response to low strains.
Jamming in knitted fabrics is qualitatively different from jamming in granular systems in a multitude of ways.
For one, knitted fabrics are far from amorphous, as they typically possess periodic translational symmetry.
Unlike jammed, athermal granular systems, the fabrics show jammed behavior under tension and move from a jammed state to a softened state (Fig.~\ref{fig:f1}) with greater applied force~\cite{Postle2002}.
Here, we argue that the change in mechanical properties due to yarn confinement in tightly-knitted fabrics can be regarded as analogous to the jamming behavior seen in other athermal, granular systems.

When an external force is applied to unjammed knitted fabrics, there is a wide linear elastic regime (extending anywhere from $\sim$10\% to $\sim$300\% strain) before the gradual onset of strain-stiffening behavior becomes evident~\cite{htoo_3-dimension_2017, Poincloux2018PRX, Duhovic2006, Abel2012,singal_programming_2024}.
The linear elastic moduli and strain range are highly dependent on stitch pattering and yarn properties~\cite{singal_programming_2024, ding_unravelling_2024}.
Knitted fabrics have a slip-knot microstructure that allow the constituent yarn to change in shape, giving rise to a highly extensible response~\cite{kyosev_3d_2005, wadekar_fea_2020} compared to woven, felted, and other non-woven fabrics.
Because of this, knitted fabrics are ideal for wearable electronics \cite{hu_review_2012}, soft robotics \cite{sanchez_3d_2023}, and medical applications \cite{Magnan2020}.

However, in highly dense knits (which can be manufactured by machine by applying additional tension to the yarn during knitting or by hand by using needles that are small compared to the yarn radius), a much stiffer linear regime develops for low strain.
This low-stress high-rigidity regime then softens into low stiffness behavior at intermediate strain (Fig.~\ref{fig:f1}b).
Other work has recorded this behavior in experimental stress-strain data and either attributed it to frictional effects \cite{Duhovic2006} or ignored it completely.
This low-strain-high-modulus behavior has been seen in fabrics made of both compressible \cite{luan_auxetic_2020, amanatides_characterizing_2022, singal_programming_2024} and incompressible yarns \cite{Poincloux2018PRX, poincloux_crackling_2018, Duhovic2006}.
For fabrics used in technological implementations such as strain sensors \cite{Seyedin2019}, a jammed response can limit the sensitivity of the sensor.

\begin{figure*}
    \vspace{-.5pt}
    \centering
    \includegraphics[width = 17.8cm]{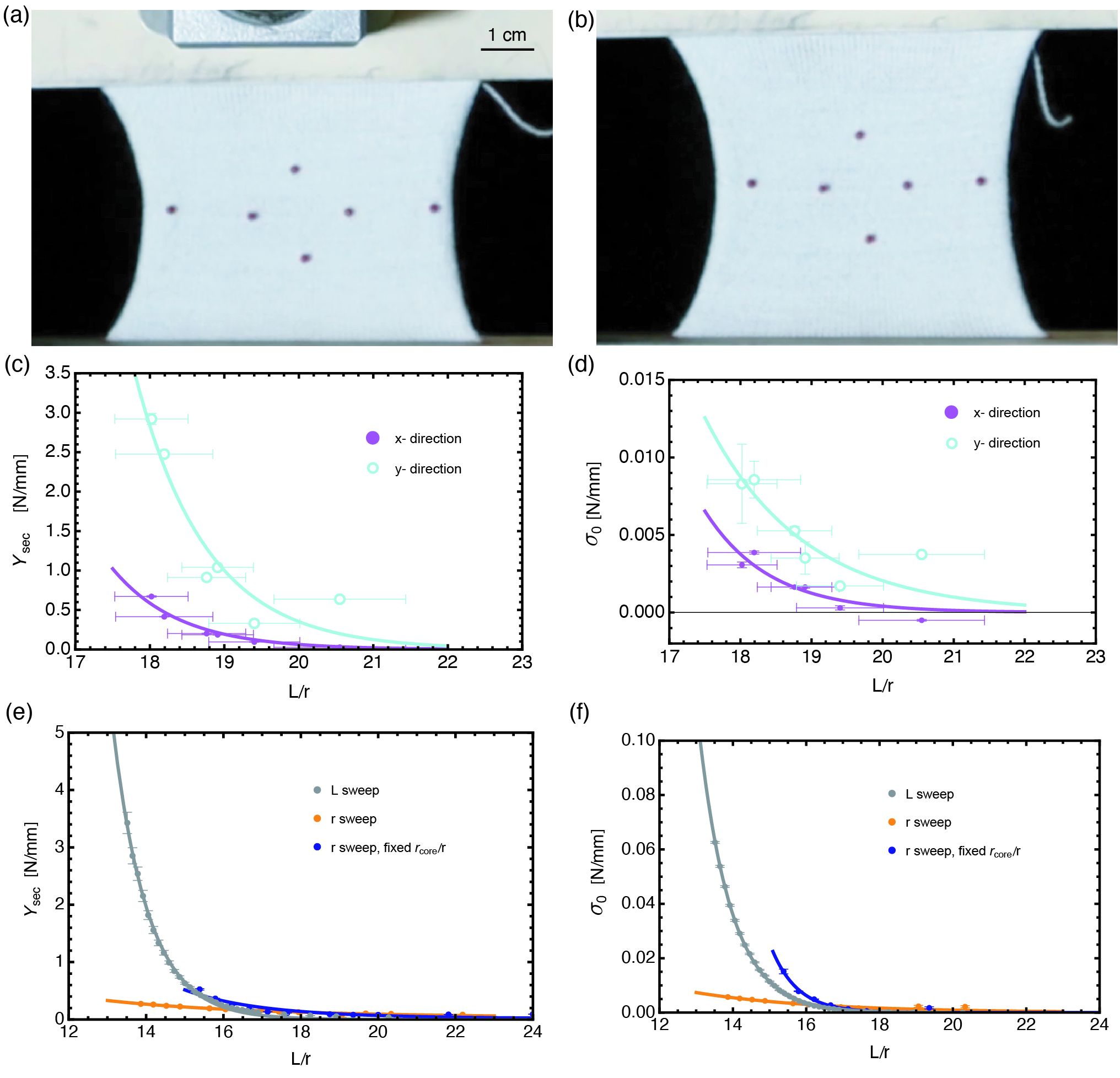}
    \caption{
    (a) Photograph of an experimental sample made at the smallest sampled ``stitch size'' setting (10.5) before it is stretched by the UTM.
    (b) Photograph of the same sample at the maximum stretch with an applied force of 60 N.
    The stress-strain plots for all experimental samples can be found in Fig. \textcolor{black}{S1}.
    Photo credit: A. Patrick Cachine.
    (c) Experimental measurements of the secant modulus $Y_{\rm sec}$ for fabrics of different ``stitch size'' settings stretched in the $x$- (purple) and $y$-directions (blue).
    (d) Experimental measurements of the offset stress $\sigma_0$ for fabrics of different ``stitch size'' settings stretched in the $x$- and $y$-directions (purple and blue, respectively).
    The error bars in the  experimentally measured $L/r$ values are standard errors of the mean, discussed in Supplementary Section \textcolor{black}{II} and shown in Fig. \textcolor{black}{S2}.
    (e) Values for the secant modulus $Y_{\rm sec}$ for simulation sweeps of the length of yarn per stitch $L$ (gray), the yarn radius $r$ (orange), and yarn radius for fixed ratio of $r_{\rm core}/r$ (dark blue) stretched in the $x$-direction.
    (f) Values for the offset stress $\sigma_0$ for simulation sweeps of the length of yarn per stitch $L$ (gray), the yarn radius $r$ (orange), and yarn radius for fixed ratio of $r_{\rm core}/r$ (dark blue) stretched in the $x$-direction.
    The yarn radius sweeps had a length of yarn per stitch $L=12.2~{\rm mm}$.
    The yarn radius for fixed ratio of $r_{\rm core}/r$ had a length of yarn per stitch $L=12.0~{\rm mm}$.
    All other simulation parameters are as described in \ref{tab:simparams}.
    Results for fabrics stretched in the y-direction are similar and shown in Fig. \textcolor{black}{S4}.
    Equations for the lines of best fit for c-f can be found in Table \textcolor{black}{S1}.
    }
    \label{fig:f2}
\end{figure*}

Here, we examine the macroscopic response of knitted fabrics, both experimentally and computationally, to determine a classification for \emph{jammed}, \emph{softened}, and \emph{unjammed} regimes as we change the applied stress and the ``density'' -- as approximated by the dimensionless parameter $L/r$, where the length of yarn per stitch $L$ is normalized by yarn radius $r$.
We use simulations to determine the micromechanical origins of this jammed behavior, and we determine that in the jammed regime individual stitches are geometrically confined such that the yarn must rearrange in order to accommodate applied stress.
In these densely-knitted fabrics, the stitches are highly compressed, which restricts the ability of the yarn to change shape to accommodate stretching.
This results in a stiffer response at low strain.
The interplay between yarn bending and the change in contacts between segments of yarn within a stitch leads to anisotropic jamming mechanisms, where geometric changes mediated by bending stiffness cause jamming in the $x$-direction and changing contacts due to compression causes jamming in the $y$-direction.
This anisotropic jamming mechanism can be seen in the aspect ratio of the fabric, the energy of individual stitches, and the forces generated by yarn contacts.

\section{Experimental Observations of Jamming}

We fabricated experimental samples of stockinette fabric made from lace-weight yarn on a STOLL industrial knitting machine, using the fabrication and methods outlined in \cite{singal_programming_2024} and additionally described in Supplementary Section~\textcolor{black}{I}. 
Each sample was made of the same yarn with identical manufacturing parameters except for the ``stitch size," which is a dimensionless parameter within the M1 Plus software that effectively changes the yarn tension of the machine.
Changing the yarn tension leads to changes in the length of yarn per stitch \emph{and} yarn radius within the samples, as described in Supplementary Section~\textcolor{black}{II}.
Both the length of yarn per stitch and yarn radius decrease as the ``stitch size" parameter decreases, as seen in Fig ~\textcolor{black}{S2}.
Images of experimental samples being stretched can be seen in Fig.~\ref{fig:f2}a,b.
Stress-strain results for these fabrics can be seen in Fig.~\textcolor{black}{S1}.

Jamming can be found in experimental samples for multiple different ``stitch size" settings (Fig.~\textcolor{black}{S1}).
As the ``stitch size'' decreases, the fabrics become tighter (Fig.~\ref{fig:f5}c-d), stiffer, and begin to show jammed behavior (Fig.~\textcolor{black}{S1}).
All fabrics that display jammed behavior in the $y$-direction also show it in the $x$-direction, but the converse does not hold.
Since both the length of yarn per stitch and yarn radius change in these samples (Fig.~\textcolor{black}{S2}), one or more of these changes leads to changes in jamming.

\section{Modeling Micromechanics}

To investigate the micro-mechanical details of the jammed regime, we utilized the simulation framework described in~\cite{singal_programming_2024}.
We include a brief overview here.
These simulations model a single unit cell of the knitted fabric.
For stockinette fabric, the unit cell is a single knit stitch, shown in Fig.~\ref{fig:f1}a.
In our model, two main types of deformation to the yarn cost energy: bending and compression.
Yarn bending is modeled as Euler elastica, such that the total bending energy $E_b$ is given by
\begin{equation}
    E_{b} = \frac{B}{2} \int_0^L {\rm d}s\, \kappa^2(s)\,\,, 
\end{equation}
where $B$ is the bending modulus, $L$ is the total length of yarn within a stitch, and $\kappa$ is the curvature of the yarn.

The compression force experienced when two non-neighboring segments of yarn come into contact is modeled as an inverse power law for a soft-shell, hard-core yarn~\cite{kaldor8, sperl_estimation_2022, singal_programming_2024, ding_unravelling_2024} in terms of a non-dimensional measure of the distance that the yarn is compressed, $\zeta(s,s') \equiv \left(R(s,s') - 2r_{\rm core}\right)/(2r - 2r_{\rm core}),$ where $r$ is the yarn radius, $r_{\rm core}$ is radius of the incompressible inner core, and $R(s,s')$ is the straight-line separation between two segments of yarn that are parameterized by arclengths $s$ and $s'$, respectively.
These quantities are diagrammed in Fig.~\textcolor{black}{S3}.
For $\zeta < 1$, the yarn is compressed, and the compression energy is given by:
\begin{widetext}
\begin{equation}
    E_{c}(\zeta) = k\frac{(2r - 2r_{\rm core})^2}{p(p-1)} \iint {\rm d}s\,{\rm d}s'
    \left[\zeta^{1-p}(s,s') - 1 - (p-1)\left(1 - \zeta(s,s')\right)\right]\,\,,
    \label{eq:compenergy}
\end{equation}
\end{widetext}
where the compression constant $k$ sets the energy scale, $p$ is the exponent characterizing the nonlinearity of the force law, and the double integral is taken over all yarn segments.
Both the compression constant $k$ and characteristic exponent $p$ are fit to experimental compression data~\cite{singal_programming_2024}.
The total yarn energy is thus the sum of bending and compression energies, $E_{\rm total} = E_{b} + E_{c}$.

The simulation takes an initial configuration of yarn with correct stitch topology and then relaxes its shape to find the minimum energy configuration, while constraining the length of yarn in the simulation box to be fixed.
The yarn center-line is modeled as a $\mathcal{C}^2$ continuous curve through a B\'ezier spline representation using degree-5 Bernstein polynomials (see~\cite{singal_programming_2024} for more details).
Unless otherwise specified, the simulation inputs are as described in Table~\ref{tab:simparams}.
To generate a stress-strain curve from the simulation, many simulations are run for different unit cell dimensions to represent the fabric being pulled in one direction.
Unlike the simulation method in~\cite{singal_programming_2024}, the zero-force configuration is found by running the simulation with only the length constraint.
Instead of sweeping over one of the stitch cell dimensions and interpolating to predict the zero-force configuration, both dimensions of the unit cell are allowed to vary freely to find the global minimum energy state of the fabric.
From that zero-force state, we then fix the $x$- (or $y$-) dimension of the unit cell, allowing the other dimension to vary freely and the yarn shape to change.
We iteratively repeat this process with a larger and larger $x$- (or $y$-) dimension until we reach a stress of 0.4~N/mm.
Notably, these simulations solve for a minimum energy and do not incorporate yarn friction.
In previous results with a similar simulation framework, these static simulations are able to generate jammed fabric behavior with the characteristic high modulus at low strain (see Figure 2 in~\cite{singal_programming_2024}).

\section{Reducing or Inducing Jamming}

To understand how changing yarn properties and manufacturing parameters affect jamming, we must first quantify fabric jamming.
We identify two different metrics for quantifying the behavior around the jamming transition that can be measured from any stress-strain curve.
First is the slope of the jammed part of the curve at low strain, which we estimate with the secant modulus:
\begin{equation}
    Y_{\rm sec} = \frac{\sigma}{\epsilon}
\end{equation}
where $\sigma$ is the stress measured at a set value of linear strain $\epsilon$.
For uniformity, we have fixed the strain by which we measure $Y_{\rm sec}$ to $\epsilon = 0.02$ for both experimental and simulation results.
This small-strain cut-off is large enough to capture sufficient data for a determination of $Y_{\rm sec}$ for simulations yet small enough to describe most small-scale jamming in experimental results.

An alternate measure is the applied stress needed bring the fabric out of its jammed regime.
One estimate is found by linearly extrapolating the post-jammed stress-strain curve to the $\epsilon = 0$ limit, $\sigma_0$, which is the $\sigma$-axis intercept of the linear fit to the post-jammed regime and represents the offset stress.
For fabrics with large jammed regimes and relatively small values of maximum strain, $\sigma_0$ becomes difficult to define accurately (see Supplementary Section~\textcolor{black}{IV}, Figs.~\textcolor{black}{S7 and S8}).
This limitation is particularly significant for fabrics stretched in the $y$-direction, which typically have larger jammed regimes and are generally less extensible (shown in Fig.~\textcolor{black}{S1}).
These two metrics from the stress-strain relation can be applied to both experimental and simulated data.
Both quantities are relative; with a set of $Y_{\rm sec}$ (or $\sigma_0$) quantities, the data can be qualitatively separated into jammed and unjammed.

\begin{table}[t]
    \centering
    \begin{tabular}{|c|c|}
        \hline
       Quantity  & Input \\
       \hline
       \hline
        $B$ & 90 mN mm\textsuperscript{2} \\
        $k$ & 0.6 mN mm\textsuperscript{-2} \\
        $r$ & 0.74 mm \\
        $r_{core}$ & 0.325 mm \\
        $p$ & 2.4 \\
        \hline
    \end{tabular}
    \caption{Simulation inputs for bending modulus ($B$), compression scaling constant ($k$), yarn radius ($r$), core radius ($r_{core}$), and power in the compression potential ($p$).
    Unless otherwise specified, these are the simulation inputs for all simulations.
    These were taken from experimental values in \cite{singal_programming_2024}.
    }
    \label{tab:simparams}
\end{table}

Next, we examine conditions under which fabrics become jammed.
In analogy with the jamming phase diagram for athermal granular systems~\cite{behringer_physics_2019}, we propose that fabric jamming occurs for sufficiently high stitch density (which occurs when the amount of yarn forming the stitch is small relative to the yarn radius) and low enough applied stress, as shown in Fig.~\ref{fig:f1}c.
Notably, the length of yarn per stitch $L$ and yarn radius $r$ are not independent and both vary with the ``stitch size" setting on the M1 Plus software used to program our industrial knitting machine as described in Supplementary Section~\textcolor{black}{V}.

The secant modulus $Y_{\rm sec}$ and offset stress $\sigma_0$ were calculated for all experimental fabric samples in both directions, shown in Fig.~\ref{fig:f2}c,d.
Generally, increasing $L/r$ (``stitch size") leads to less jammed fabrics as seen by decreasing secant modulus $Y_{\rm sec}$ and offset stress $\sigma_0$.
This is consistent with qualitative examination of the stress-strain curves in Fig.~\textcolor{black}{S1}.
We can see that the fabrics have larger $Y_{\rm sec}$ in the $y$-direction than the $x$-direction, implying that they are ``more jammed" or harder to soften in the $y$-direction as seen in Fig.~\ref{fig:f2}c.
Fits for this data, and all other fits contained in Fig.~\ref{fig:f2}, can be found in Table~\textcolor{black}{S1}.

We measure the response of the jamming metrics defined above for a variety of simulated stitches.
Since both the $x$- and $y$-directions are similarly affected by length of yarn per stitch and yarn radius, we display results for the $x$-direction in Fig.~\ref{fig:f2}e,f and the $y$-direction in Fig.~\textcolor{black}{S4}.
As the length of yarn per stitch $L$ increases, both the secant modulus $Y_{\rm sec}$ and the offset stress $\sigma_0$ decrease nonlinearly.
The secant modulus of the jammed regime decreases until it converges with the Young's modulus of the unjammed fabric, indicating that the fabric has reached the un-jammed regime.
The offset stress goes to zero and the exact jamming transition is not readily found in this data without establishing a cut-off value.
The effects of other parameters, such as yarn compressive stiffness, on these jamming metrics can be found in Supplementary Section~\textcolor{black}{III} and Fig.~\textcolor{black}{S5}.

\begin{figure*}[t!]
    \centering
    \includegraphics[width = 17.2cm]{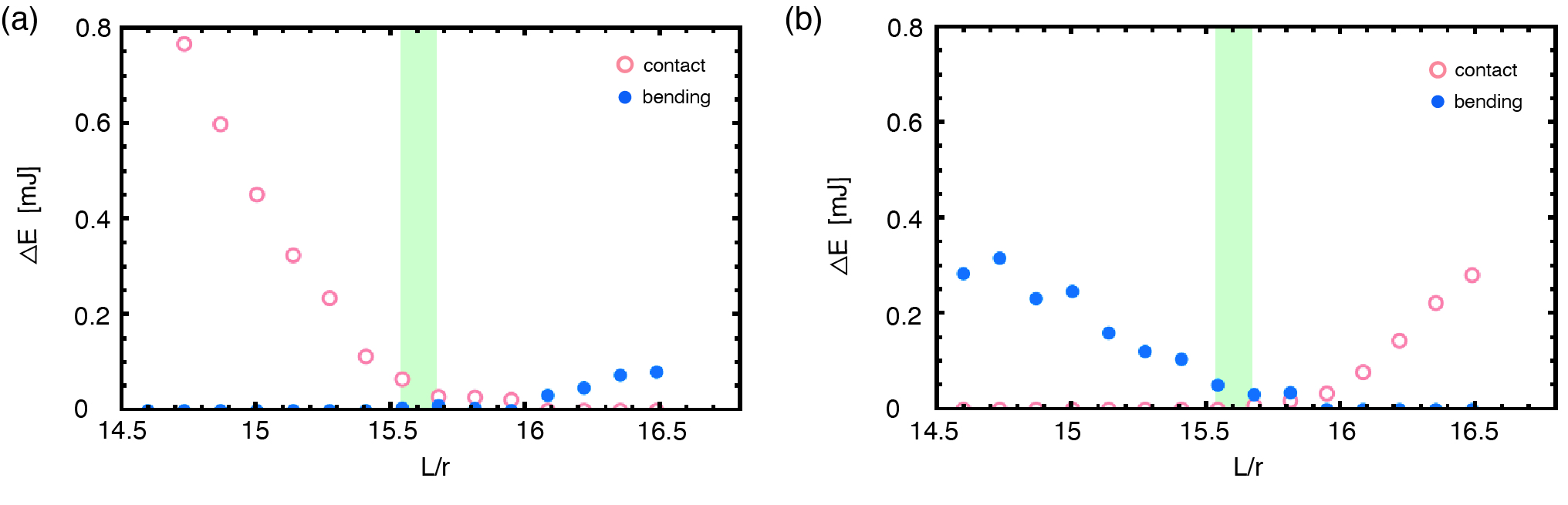}
    \caption{Energy gap for the contact (coral) and bending (blue) energies as a function of length of yarn per stitch for compression scaling constant $k=0.1$ mN/mm\textsuperscript{2} pulled in the $x$-direction (a) and the $y$-direction (b).
    The light green region shows where in length of yarn per stitch parameter space the fabric moves from jammed ($\Delta E_c > 5*10^{-5}$ J) to un-jammed ($\Delta E_c < 5*10^{-5}$ J).
    Both log-linear and rest-energy rescaling depictions of these plots are given in Fig. \textcolor{black}{S15}.}
    \label{fig:f3}
\end{figure*}

Changing the yarn radius alone has a much smaller effect on the two jamming metrics (Fig.~\ref{fig:f2}e,f).
Although increasing the yarn radius leads to increasingly-jammed fabrics, as seen by an increase in both $Y_{\rm sec}$ and $\sigma_0$, the response is distinct from changes to $L$, suggesting that additional parameters are needed to characterize the jamming diagram.
Since the exact way the core radius changes with ``stitch size" is unknown, we also varied the yarn radius with a fixed ratio of $r_{\rm core}/r = 0.45$.
In simulations, the ratio $r_{\rm core}/r$ seems to have a larger effect on the jamming behavior than changing the yarn radius $r$ alone (seen in Fig.~\ref{fig:f2}e,f and Fig.~\textcolor{black}{S4}) and the stiff jammed regime is larger as measured by both the secant modulus and offset stress.
We can rationalize the augmented jamming response to changes in $r_{\rm core}$ in relation to the other yarn lengthscales given that the compression stiffness has a nonlinear dependence on $r_{\rm core}$.
It is possible that the fixed ratio of core radius to yarn radius may induce more jamming with a larger secant modulus $Y_{\rm sec}$ than the length of yarn per stitch, but these simulations for large $r$ (small $L/r$) became numerically unstable.

The trends exhibited by the simulation results are consistent with the trends seen in experimental data.
Comparison between the fits listed in Table \textcolor{black}{S1} and visual examination indicate that both the experimental $Y_{\rm sec}$ and $\sigma_0$ more closely resemble simulation results for changing the length of yarn per stitch and changing yarn radius with fixed ratio of $r_{\rm core}/r$ and not changing yarn radius alone.
Simulation results for the length of yarn per stitch and yarn radius sweeps support experimental trends of larger secant modulus $Y_{\rm sec}$ in the $y$-direction than the $x$-direction (Fig. \ref{fig:f2}e,f and Fig.~\textcolor{black}{S4}).

\section{Energy Markers for Jamming}

When conducting simulations, we have additional information about the energy profile that is currently unavailable in experiments.
Since the energy distribution between bending and compression energy is unknown during uniaxial experiments, simulations can provide critical insights to markers and mechanisms behind jamming.
As a way of measuring how jammed a fabric is, we use the energy gap
\begin{equation}
    \Delta E_i = E_{i,{\rm rest}} - E_{i,{\rm min}},
\end{equation}
\noindent
where $E_{i,{\rm rest}}$ is one component of the energy at the zero-force configuration and $E_{i,{\rm min}}$ is the minimum value of that same energy component as the fabric is being stretched.
As described in simulation methods, the two components of the energy are bending and compression such that $i \in \{b, c\}$.

\begin{figure*}[t]
    \centering
    \includegraphics[width = 17.8cm]{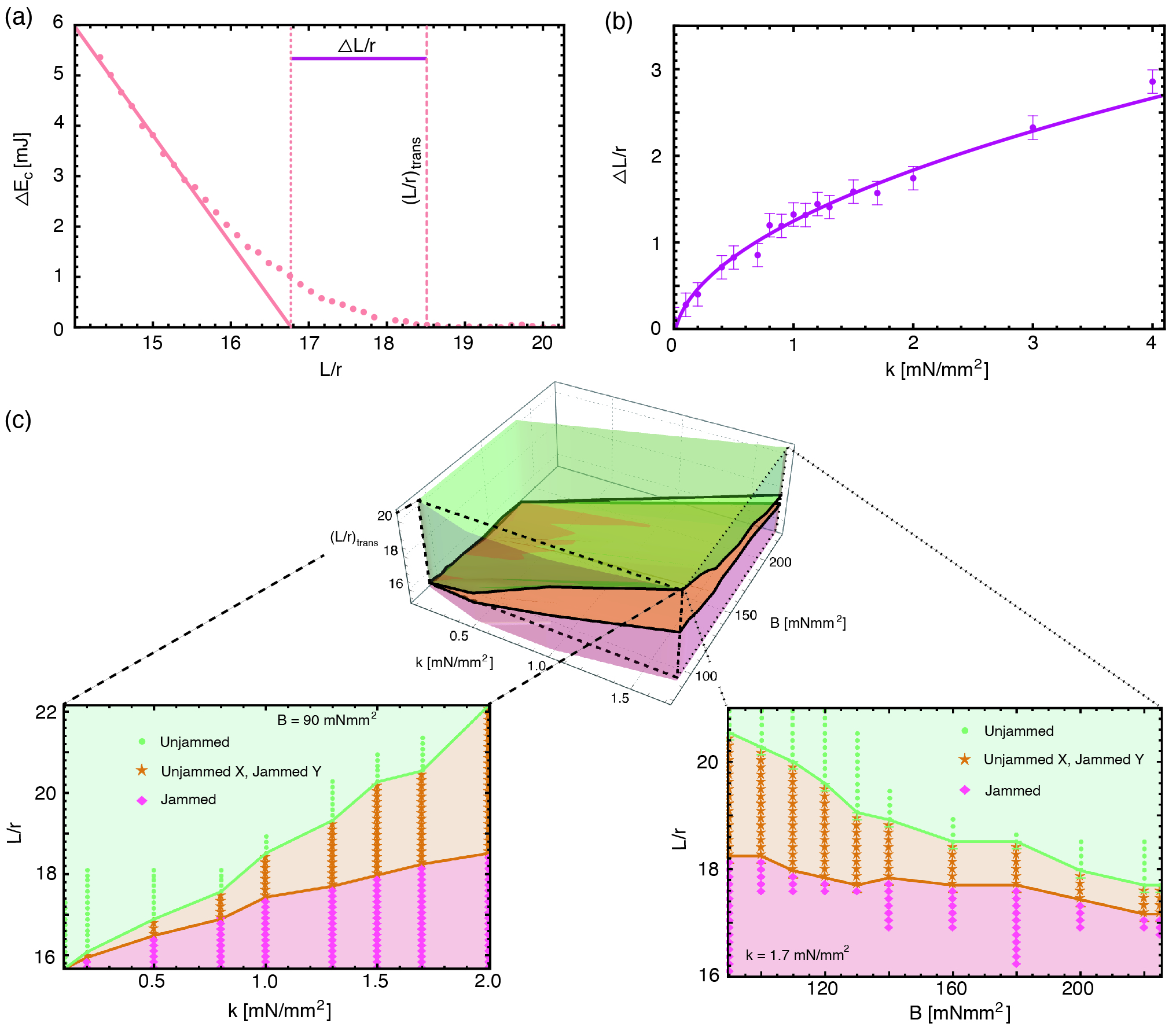}
    \caption{(a) Contact energy gap as a function of length of yarn per stitch for compression scaling constant $k=2~{\rm mN/mm^2}$ for fabrics stretched in the $x$-direction.
    The onset of un-jamming is marked with the vertical dashed line labelled $(L/r)_{\rm trans}$.
    $\Delta L/r$ is a measure of the size of the jamming transition and defined as the distance between the x-intercept of the linear fit to the jammed regime (the dotted line) and $(L/r)_{\rm trans}$.
    The linear fit is described in Table \textcolor{black}{S4}.
    Error in the energy calculation is described in Supplementary Section \textcolor{black}{VII} and Table \textcolor{black}{S2}.
    (b) The jamming transition region $\Delta L /r$ as a function of the compression scaling constant $k$.
    The jamming transition region becomes larger as $k$ increases and the yarn moves further from the incompressible limit, $k\rightarrow 0$.
    Error in $\Delta L/r$ is described in Fig. \ref{fig:f3}.
    (c) Phase diagram showing the jammed and unjammed regimes for different lengths of yarn per stitch $L$, compression scaling constants $k$, and bending moduli $B$.
    The top center plot is a phase space diagram showing $(L/r)_{\rm trans}$ as a function of the compression scaling constant $k$ and the bending modulus $B$ for the $x$-direction (orange) and the $y$-direction (green).
    The lower left shows a slice of this 3D phase space with varying $k$ and the lower right is a slice of varying $B$.
    The pink regimes are jammed in both the $x$- and $y$-directions, the orange regimes are jammed only in the $y$-direction, and the green regimes are unjammed.
    The onset of unjamming, $(L/r)_{\rm trans}$, is given by the solid lines, where an orange solid line is the end of jamming in the $x$-direction and a green solid line is the end of jamming in the $y$-direction.
    Increasing $k$ leads to a wider gap of $(L/r)_{\rm trans}$ between the $x$- and $y$-directions; whereas, increasing $B$ closes that gap.
    }
    \label{fig:f4}
\end{figure*}

Changing the length of yarn per stitch $L$ for different values of the compression scaling constant $k$ indicates the presence of three regimes within the energy difference: jamming, a transition region, and un-jammed fabric.
For extension in both the $x$- and $y$-directions, we see all three regimes, though the energy contributions differ for each direction.
Jamming in the $x$-direction has an energy gap in the contact energy whereas jamming in the $y$-direction has an energy gap in the bending energy (Fig.~\ref{fig:f3}).
The linear regime of the curve in Fig.~\ref{fig:f4}a indicates jamming when the energy-minimizing configuration of the fabric does \emph{not} correspond to the minimum contact energy (for the $x$-direction) or the minimum bending energy (for the $y$-direction).
The horizontal region along the $x$-axis indicates an unjammed or nearly unjammed state, when the energy-minimizing configuration also corresponds the minimum contact energy (for the $x$-direction) or the minimum bending energy (for the $y$-direction).
We describe the onset of unjamming, the first stitch length with an energy gap below our error threshold, as $(L/r)_{\rm trans}$ (Supplementary Section~\textcolor{black}{VII}).
We characterize the distance between the $x$-intercept of the linear jammed regime and $(L/r)_{\rm trans}$ as $\Delta L/r$.
$\Delta L/r$ shows that the transition region from jammed to unjammed behavior is small for low compression constants and becomes larger for larger compression constants, shown in Fig.~\ref{fig:f4}b.

The exact jamming transition lies within this region.
Small compression constants, $k \lesssim 0.1 {\rm mN/mm^2}$, are near one of the incompressible limits, $k \to 0$, where the compression force in the outer core of the core-shell model becomes zero and the inner core remains incompressible.
At large $k$, it becomes difficult to determine if fabrics near the jamming transition region, as defined by the energy gap, are jammed or un-jammed from the stress-strain plots.

For jammed fabrics stretched in the $x$-direction, an energy gap in the contact energy implies that the bending energy is at a minimum in these jammed configuration.
The energy gap of bending energy, $\Delta E_b$, is indeed zero in the jammed regime and becomes non-zero when the contact energy gaps goes to zero for unjammed fabric configurations (Fig.~\ref{fig:f3}a).
This is evidence that yarn bending determines the behavior of the jammed regime when the fabric is pulled in the $x$-direction.
Stretching the fabric in the $x$-direction is very different from stretching in the $y$-direction due to the anisotropy of the stitches themselves (seen in Fig.~\ref{fig:f1}a), and this is reflected in the energy gap.

\begin{figure*}
    \centering
    \includegraphics[width = 17.8cm]{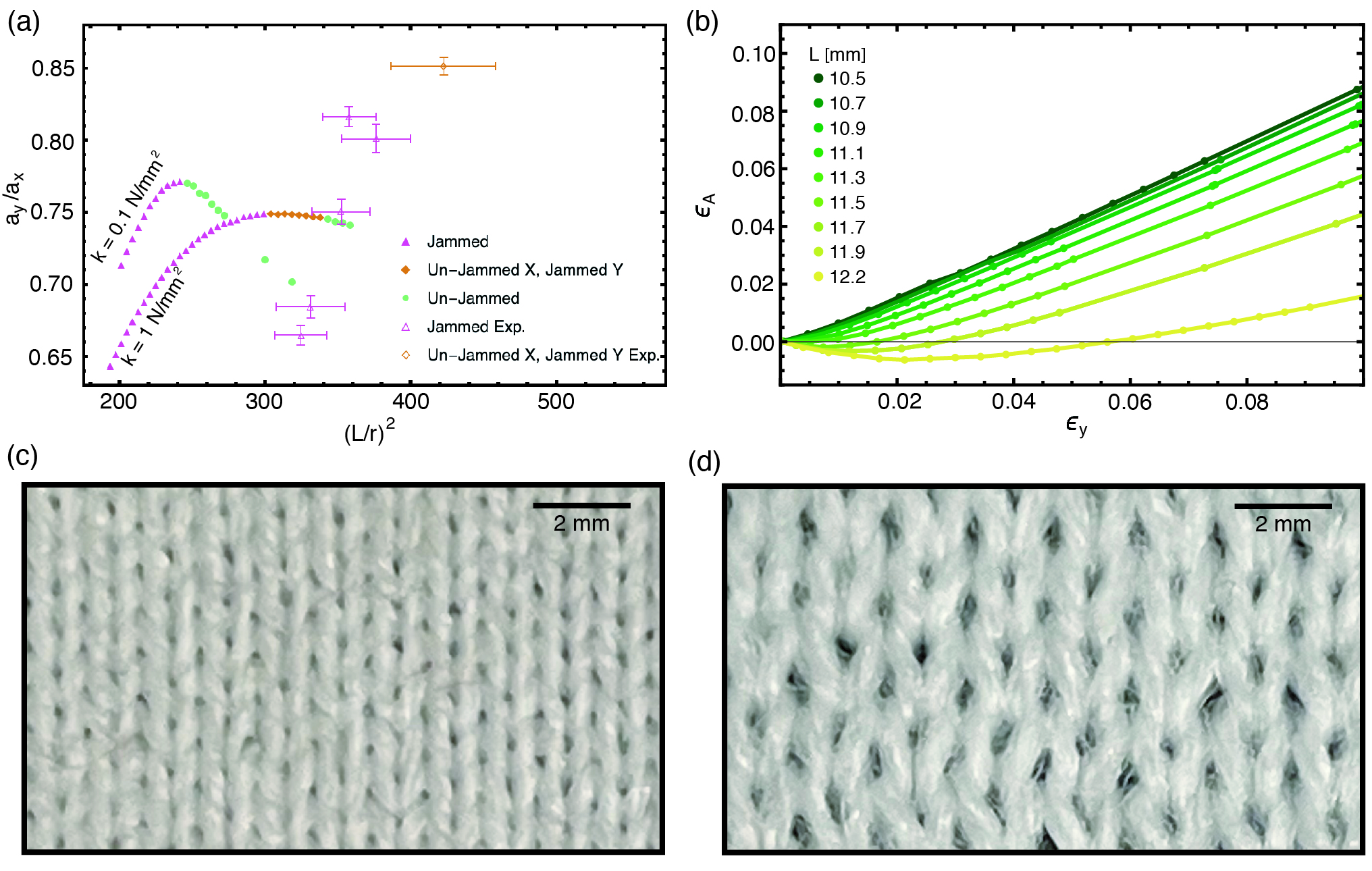}
    \caption{(a)
    Aspect ratio of a stitch $a_y/a_x$ as a function of $(L/r)^2$ for two simulation runs ($k=0.1~{\rm N/mm^2}$ and $k=1.0~{\rm N/mm^2}$) in closed symbols and experimental data in open symbols.
    For the simulation data, there are three regimes: fabrics jammed in both extension directions (pink), fabrics jammed only in the y-direction (orange), and fabrics that are not jammed in either direction (green).
    The transition from jammed in both directions to jammed only in the $y$-direction gives the peak of the aspect ratio curve.
    The experimental data is not inconsistent with this trend; though the data is sparse and the error bars are large, this transition is expected between $375 < (L/r)^2 \leq 423$ in the experimental data from the stress-strain plots in Fig. \textcolor{black}{S1}.
    Due to manufacturing constraints, the experimental data does not display a sample that is un-jammed in both extension directions.
    (b) Simulation values for the area strain as a function of strain in the y-direction for fabrics of different lengths of yarn per stitch $L$ stretched in the $y$-direction.
    Unjammed fabrics, $L>11.6~{\rm mm}$, have non-monotonic area strain that dips below zero, indicating that the zero-force configuration is not the minimum area configuration when they are stretched in the y-direction.
    Jammed fabrics, by contrast, have a zero-force configuration that is also the minimum area configuration.
    Area strain for the $x$-direction is given in Fig. \textcolor{black}{S13}, experimental area strain is given in Fig. \textcolor{black}{S14}, and these results are discussed in more detail in Supplementary Section \textcolor{black}{VI}.
    (e) Photograph of the experimental sample made at the smallest sampled machine tension setting (10.5).
    (f) Photograph of the experimental sample made at the largest sampled ``stitch size'' setting (13).
    These photographs show how changing the ``stitch size'' setting leads to visible differences in the length of yarn per stitch, the area of the stitches, the stitch aspect ratio, and the yarn radius.
    These effects are quantified in Supplementary Section \textcolor{black}{II}.
    Photo credit: A. Patrick Cachine.
    }
    \label{fig:f5}
\end{figure*}

For jammed fabrics stretched in the $y$-direction, the change in the compression energy $\Delta E_c$ is zero, and we see that the contact energy is at a minimum (Fig.~\ref{fig:f3}b) even though the total contact energy is smaller than the total bending energy~\cite{singal_programming_2024}.
This implies that yarn contacts dominate in this regime.
For increasing compression scaling constant $k$, the onset of unjamming, $(L/r)_{\rm trans}$, moves to larger and larger values of $L$, shown in Fig.~\ref{fig:f4}c.
For small values of $k$ (the incompressible regime), the transition coincides for both the $x$- and $y$-directions.
As $k$ increases, the $y$-direction becomes ``more jammed'' than the $x$-direction.
We find that the range of $L$ values that are jammed in the $y$-direction but not jammed in the $x$-direction becomes larger as $k$ increases.

As the bending modulus increases, the onset of unjamming, $(L/r)_{\rm trans}$, remains nearly constant for fabrics stretched in the $x$-direction.
There is a slight decrease in $(L/r)_{\rm trans}$ for relatively large compression constant values ($k > 1\, {\rm mN/mm^2}$), seen in Fig.~\ref{fig:f4} and Fig.~\textcolor{black}{S6}.
By comparison, the bending modulus $B$ has a minor effect on the jamming behavior, with larger values of $B$ corresponding to jamming occurring at smaller stitch lengths $L/r$ (i.e.~denser fabrics).
This decrease in $(L/r)_{\rm trans}$ is larger for $y$-direction jamming.
As the bending stiffness of the yarn increases, the compression stiffness plays a reduced role in determining yarn geometry.
Consequently, $(L/r)_{\rm trans}$ for $x$- and $y$-direction converge to similar values for increasing $B$, similar to the $k \to 0$ limit.
As shown in Fig.~\textcolor{black}{S6}, $(L/r)_{\rm trans}$ in the $x$-direction never becomes significantly larger than $(L/r)_{\rm trans}$ in the $y$-direction, implying that fabrics that are jammed in the $x$-direction and not the $y$-direction are very unlikely to exist or impossible to manufacture.

\section{Anisotropic Markers in Fabric Dimensions}

Another anisotropic marker of jamming arises in the shape of the stitch unit cell.
For fabrics jammed in the $y$-direction, the zero-strain configuration is the minimum area-per-stitch over the entire strain sweep (Fig.~\ref{fig:f5}b).
For un-jammed fabrics, the zero-strain configuration is \textit{not} the minimum area-per-stitch; as the fabric is pulled in the $y$-direction, the area-per-stitch decreases and then increases, shown in Fig.~\ref{fig:f5}.
This indicates that fabrics jammed in the $y$-direction are geometrically confined and at the smallest possible area-per-stitch for their set of yarn and manufacturing parameters.
Fabrics stretched in the $x$-direction have a non-monotonic area-per-stitch as a function of strain regardless of whether or not the fabric is jammed in that direction and simulated fabrics all take on the minimum area-per-stitch at zero-strain (Fig.~\textcolor{black}{S13}).
We chose to analyze the stitch area instead of the stitch volume because existing experimental procedure can determine area strain.
However, experimental area strain results, calculated from the motion of internal pins as seen in Fig.~\ref{fig:f2}e,f, have very large errors that obscure possible trends in area strain due to jamming.
These results are in Fig.~\textcolor{black}{S14} and discussed in more detail in Supplementary Section~\textcolor{black}{VI}.

The aspect ratio of the stitch cell for the zero-strain configuration also gives information on jamming.
For small compression scaling constant $k$ in the in-compressible limit, the aspect ratio $a_y/a_x$ as a function of $(L/r)^2$ reaches a maximum near the transition from jammed to un-jammed (Fig.~\ref{fig:f5}a).
For larger $k$, with softer transitions and a mismatch between jamming transitions in the $x$- and $y$-directions, the aspect ratio reaches a maximum when the $x$-direction becomes un-jammed.
This is also true for other simulation results with mis-matched jamming transitions in the $x$- and $y$-directions.
Experimental measurements of the fabric aspect ratio show a similar trend (Fig.~\ref{fig:f5}a).
Due to the small number of experimental samples (six) and the large variance in the measured yarn radius, it is difficult to determine if the maximum aspect ratio aligns exactly with the onset of un-jamming in the $x$-direction.

\section{The Jamming Mechanism}

\begin{figure*}[t!]
    \centering
    \includegraphics[width = 17.8cm]{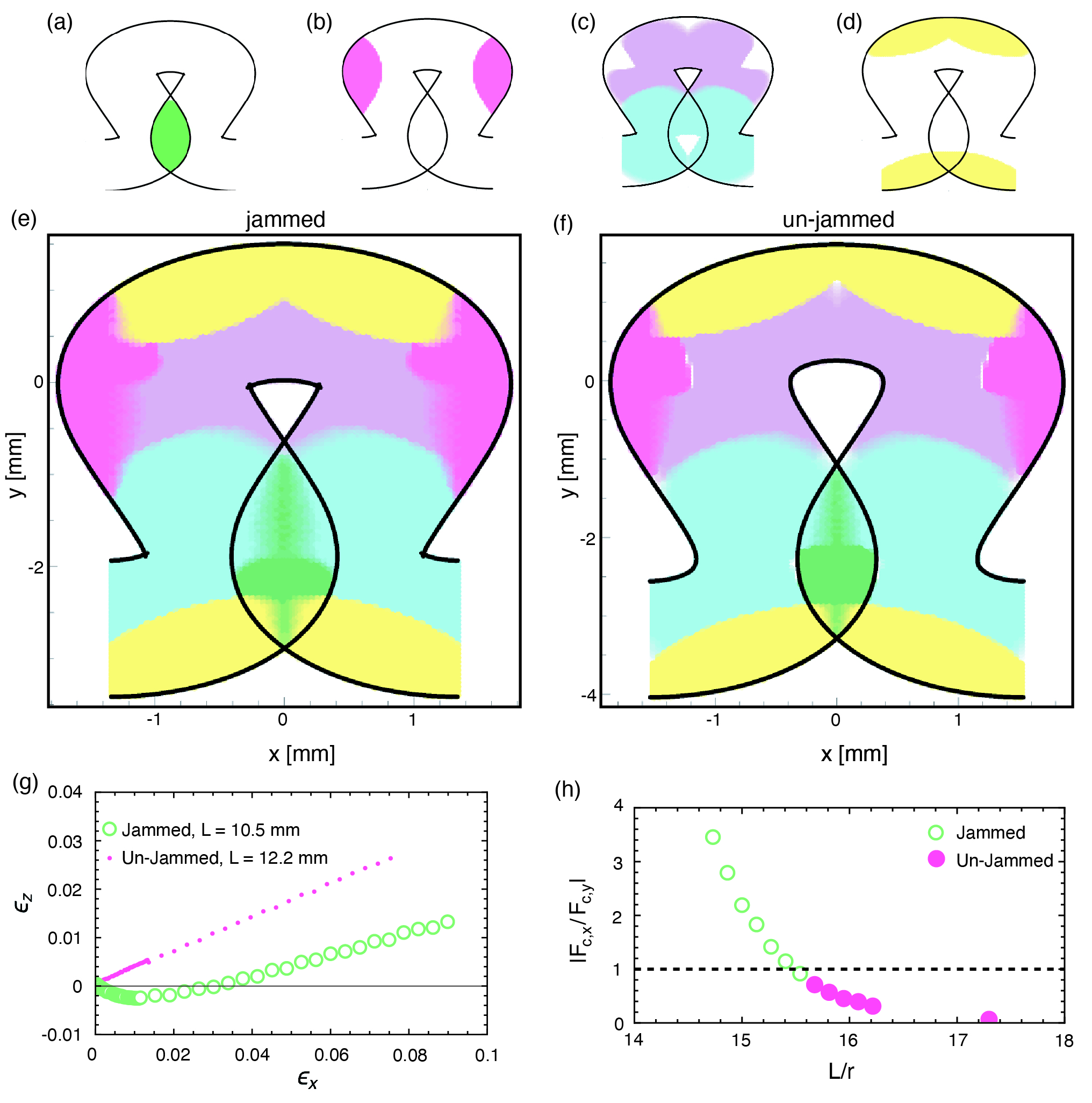}
    \caption{(a-e) Stitch contact volumes shown for a jammed fabric with a length of yarn per stitch $L=10.5~{\rm mm}$.
    (a) The self-contact shown in green.
    (b) The nearest-neighbor contact with stitches in the same row but different columns shown in magenta.
    (c) The nearest-neighbor contact with stitches in the same column a row above (lavender) or below (cyan).
    (d) The next-nearest-neighbor contact with stitches in the same column two rows above or below in yellow.
    (e) All contact volumes shown together in the stitch for a jammed fabric with length of yarn per stitch $L=10.5~{\rm mm}$.
    (f) Contact volumes shown for an un-jammed fabric with length of yarn per stitch $L=12.2~{\rm mm}$.
    The contact volumes for (a-f) look fuzzy at the edges because they are made of a semi-transparent clouds of points.
    (g) Strain in the $z$-direction as a function of strain in the $x$-direction for an un-jammed ($L=12.2~{\rm mm}$ in filled circles) and jammed ($L=10.5~{\rm mm}$ in open circles) fabric being stretched in the $x$-direction.
    The grey horizontal line is zero-strain.
    (h) Magnitude of the ratio of total contact forces in the $x$- and $y$-directions for zero-strain configurations of fabrics with different lengths of yarn per stitch $L$.
    Near the jamming transition, we see that the contact force in the $y$-direction becomes larger than the contact force in the $x$-direction.
    The dashed horizontal line represents equal contact forces in the $x$- and $y$-directions.
    Error in contact force is described in Table \textcolor{black}{S3}.
    }
    \label{fig:f6}
\end{figure*}

Using the simulations, we are able to identify four dominant types of contacts for a single stitch: (1) self-contact where yarn within a stitch is in contact with a different segment of yarn within the same stitch, (2) nearest-neighbor contact with stitches in the same row but different columns, (3) nearest-neighbor (NN) contact with stitches in the same column a row above or below, and (4) next-nearest-neighbor (NNN) contact with stitches in the same column two rows above or below.
Each of these contacts are shown individually in Fig.~\ref{fig:f6}a-d and together in Fig.~\ref{fig:f6}e,f.
We analyze the forces applied by each of the yarn-yarn contacts in Fig.~6f and Supplementary Section V (Figs.~\textcolor{black}{S9, S10, and S11}).

When stretched in the $x$-direction, jammed fabrics must first rearrange out-of-plane before they can extend.
During conventional linear elastic behavior of unjammed fabrics, the stitch enlarges in the $z$-direction.
By contrast, jammed fabrics shrink in the $z$-direction until they reach the softened regime and then expand, as shown in Fig.~\ref{fig:f6}g and Fig~\textcolor{black}{S12}.
The only contact applying non-negligible force is the NN same-column contact, but this contact is applying a force that would expand the stitch out-of-plane like unjammed fabrics, not contract as is seen in jamming (Fig.~\textcolor{black}{S11}).
Since none of the contacts are applying an out-of-plane contraction force, this indicates that the yarn bending is causing this motion, in agreement with the energy gap analysis.

\clearpage

For jamming in fabrics pulled in the $y$-direction, the energy gap data indicates that yarn contacts lead to jammed behavior.
Analyzing the individual contact areas shows that it is the the interplay between contacts that leads to changing mechanics.
For un-jammed fabrics, the contact forces in the $y$-direction dominate in the stress-free configuration; the NNN same-column contact force is the largest contributing contact force and it pushes the stitch to extend in the $y$-direction (see Fig.~\textcolor{black}{S11e}).
For jammed fabrics, the contact forces in the $x$-direction are larger than those in the extension direction.
The sum of the self contacts and same-row contacts, both of which expand the stitch in the $x$-direction (Fig.~\textcolor{black}{S11d}), become larger than the NNN same-column and NN same-column contacts in the $y$-direction, shown in Fig.~\ref{fig:f6}h.
These forces inhibit the fabric from the classical, in-plane Poisson response and prevent extension in the $y$-direction from occurring without a high enough applied force to overcome these initial, internal forces.

\begin{table}[h!]
    \centering
    \begin{tabular}{|c|c|c|}
        \hline
       contact  &  $x$-direction & $y$-direction \\
       \hline
       \hline
       self contact & extend & - \\
       same-row contact & extend & - \\
       NN same-column contact & - & compress \\
       NNN same-column contact & - & extend \\
       \hline
    \end{tabular}
    \caption{Summary of the direction of the contact force at each contact area at zero strain.}
    \label{tab:contactforcedirection}
\end{table}

This behavior is unlike the jamming mechanism in the $x$-direction; where out-of-plane deformation was required and inhibited the fabric from stretching in the $x$-direction, the contact forces themselves inhibit stretching in the $y$-direction.
As seen in Fig.~\textcolor{black}{S12b}, there is little difference between the out-of-plane low-strain response for jammed and un-jammed fabrics when pulled in the $y$-direction.
Despite these differing mechanisms, a changing number of contacts via a coordination number is not sufficient to characterize jamming when the fabric is stretched in either direction.
As seen in Fig.~\ref{fig:f6}e,f, jammed fabrics are nearly indistinguishable if we look at the contact volumes alone.

\section{Conclusion}

Knitted fabric jamming corresponds to an anomalously-stiff regime of fabric response at low strain that can be seen in experimental samples.
As we decrease the ``stitch size" in knitted fabric manufacturing, fabrics become more jammed as described by the secant modulus $Y_{\rm sec}$ and the offset stress $\sigma_0$, two metrics we have established to quantify jamming.
Using these metrics, we compare experimental results to simulated changes to the length of yarn per stitch and yarn radius.
The simulation results are qualitatively consistent with experiment and agree that fabrics are ``more jammed" in the $y$-direction.
These geometrically-confined jammed states are energy minima for the fabrics, but are generally not contact energy minima (for fabrics jammed in the $x$-direction) or bending energy minima (for fabrics jammed in the $y$-direction).
This indicates that these states are characterized by distinct distributions of residual stress.
The energy difference between the zero-strain state and the lowest energy for either bending or contact energy shows that there is a soft transition region that is dependent on the compressibility of the yarn.
Analyzing the shape of the fabric shows anisotropic markers where jamming in the $x$-direction can be seen in the aspect ratio of the stitches and jamming in the $y$-direction can be found in the area strain.
Though we cannot accurately measure the out-of-plane thickness of experimental samples, simulations suggest that fabrics jammed in the $x$-direction first contract in the $z$-direction before expanding, and that this behavior is not caused by yarn contacts but instead by yarn rearrangements to reduce yarn bending.
Contact forces show that all four primary contacts contribute to knitted fabric jamming in the $y$-direction by inhibiting linear elastic contraction in the $x$-direction.

\section{Data and Code availability}

Source code is available at \href{https://github.com/sabetta/jammed-knits}{https://github.com/sabetta/jammed-knits}. Data is available upon request.

\begin{acknowledgments}

We thank Martin van Hecke and Krishma Singal for useful conversations. \textbf{Funding:} SG and EAM were supported by National Science Foundation Grant No.~DMR-1847172. APC was supported in part by the Research Corporation for the Advancement of Science Cottrell Scholar Award Grant No.~CS-CSA-2020-162.

\end{acknowledgments}

\clearpage
\onecolumngrid

\part*{\centering {\large Supplementary Information for: ``Pulling apart the mechanisms that lead to jammed knitted fabrics"}
}

\setcounter{section}{0}
\setcounter{figure}{0}

\renewcommand{\thefigure}{S\arabic{figure}}

\renewcommand{\thesection}{S\arabic{section}}
\renewcommand{\thesubsection}{S\Alph{subsection}}

\section{Experimental Stress-Strain Results}

Each fabric was manufactured on a STOLL industrial knitting machine using the proprietary M1 Plus software and are 64 stitches wide and 64 stitches tall without buffers.
The yarn is Tamm Petit 2/30 T4201 White 100\% acrylic yarn from The Knit Knack Shop\texttrademark{}.
The samples are stockinette fabric, also known as jersey or plain knit.
Samples were made at different ``stitch size" settings, 10.5-13.
Smaller and larger ``stitch size" settings were not able to produce consistent samples without holes or machine error.
For each individual stretching direction, a sample was made with an additional buffer region in the direction of extension.
The buffer region was clamped during extension to ensure that the entire sample is free to extend during the experimental procedure.

To stretch the fabrics, we used an Instron Universal Testing Machine (UTM) Model 68SC-1 with custom 3D printed clamps with teeth to prevent fabric slipping.
These clamps provided uniform force to the entire fabric boundary.
We determined the uniaxial response by tracking the fabric dimensions in the extension direction through the proprietary UTM software and the transverse direction through visual pin tracking.
In each sample, two sets of pins were placed in the fabric.
One set for transverse direction tracking was placed at the edge of the fabric at it's midline.
Due to the natural curling of stockinette, these pins were not placed directly at the fabric boundary, but at the furthest point that would not pierce through the curled edge.
The second set of pins was used to track the deforming area and was placed in a rectangle, approximately sized to be one third of the fabric width and height.
These pins can be seen in Main Text Fig. \textcolor{black}{2a,b}.
We used Fiji (\url{https://imagej.net/Fiji}) image processing software with the TrackMate plugin to track the pins as a function of fabric extension.

\begin{figure}[h!]
    \centering
    \includegraphics[width = 12.cm]{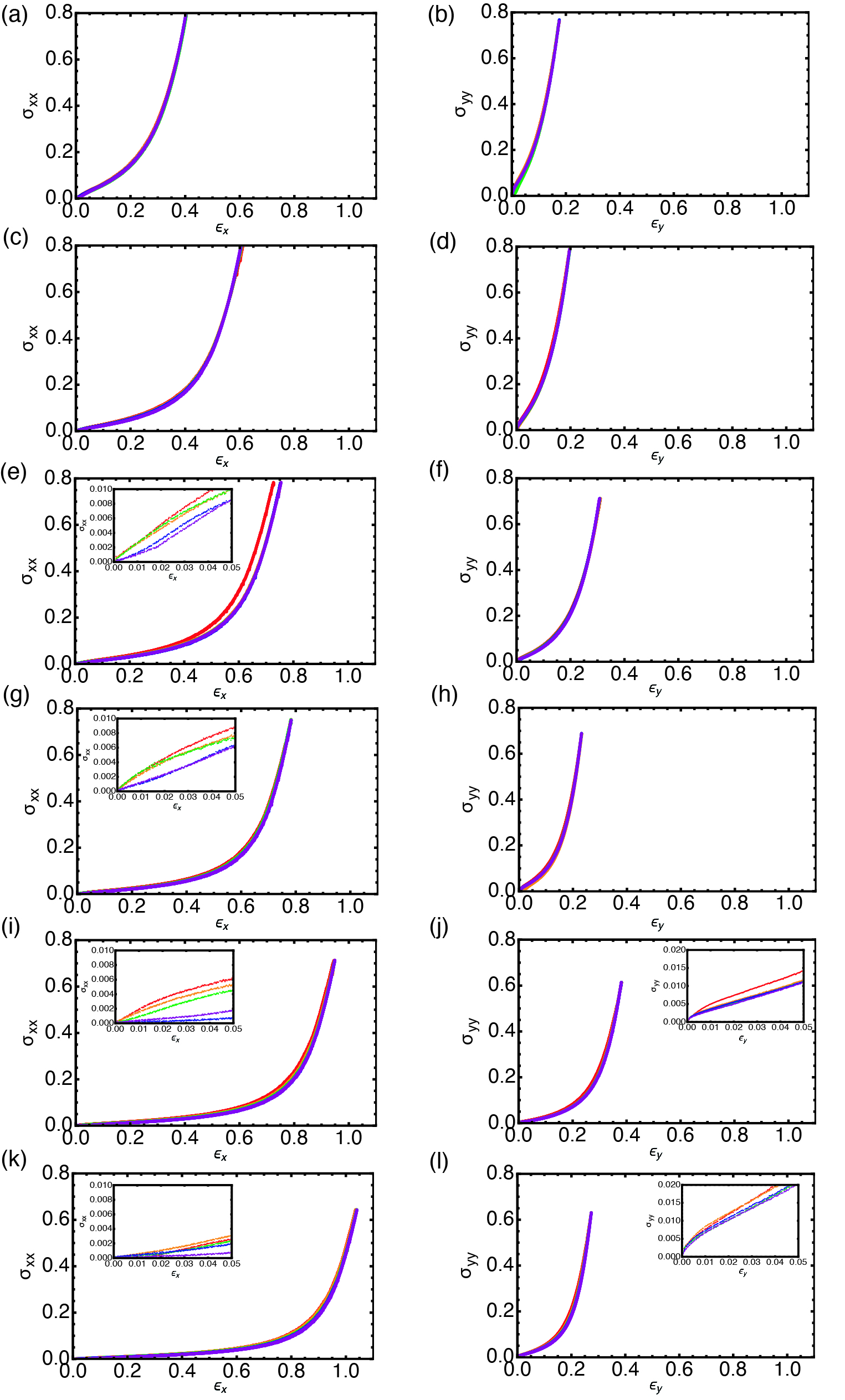}
    \caption{Experimental stress strain results for stockinette fabrics made at different machine tensions: setting 10.5 stretched in the (a) $x$-direction and (b) $y$-direction, setting 11 stretched in the (c) $x$-direction and (d) $y$-direction, setting 11.5 stretched in the (e) $x$-direction and (f) $y$-direction, setting 12 stretched in the (g) $x$-direction and (h) $y$-direction, setting 12.5 stretched in the (i) $x$-direction and (j) $y$-direction, and setting 13 stretched in the (k) $x$-direction and (l) $y$-direction.
    The color of the data represents the experimental run.
    Red is run 1, orange is run 2, green is run 3, blue is run 4, and purple is run 5.
    Insets of the small strain regime are included when jamming is not visible from the scale of the entire range of motion.}
    \label{fig:expstressstrain}
\end{figure}

Each sample was extended once without recording data to break up any fiber entanglements caused by fabric storage.
For each of the five experimental runs, the fabric was extended, the clamps returned to the rest position, the fabric was manually stretched in the transverse direction, and then the fabric rested for five minutes before the next run.
For some samples, some of the five experimental runs seem jammed while others don't (insets in Fig. \ref{fig:expstressstrain}).
This typically only occurs when the fabric is minimally jammed and the jamming is either barely or not at all noticeable on the scale of the entire extension period.
This may be a result of the intermediate process between runs.
It is possible that the intermediate transverse stretching can reduce the affects of jamming.
This phenomenon requires further study.

\section{``Stitch Size"}
\label{sec:machinetension}

\begin{figure}[h]
    \centering
    \includegraphics[width = 12.9cm]{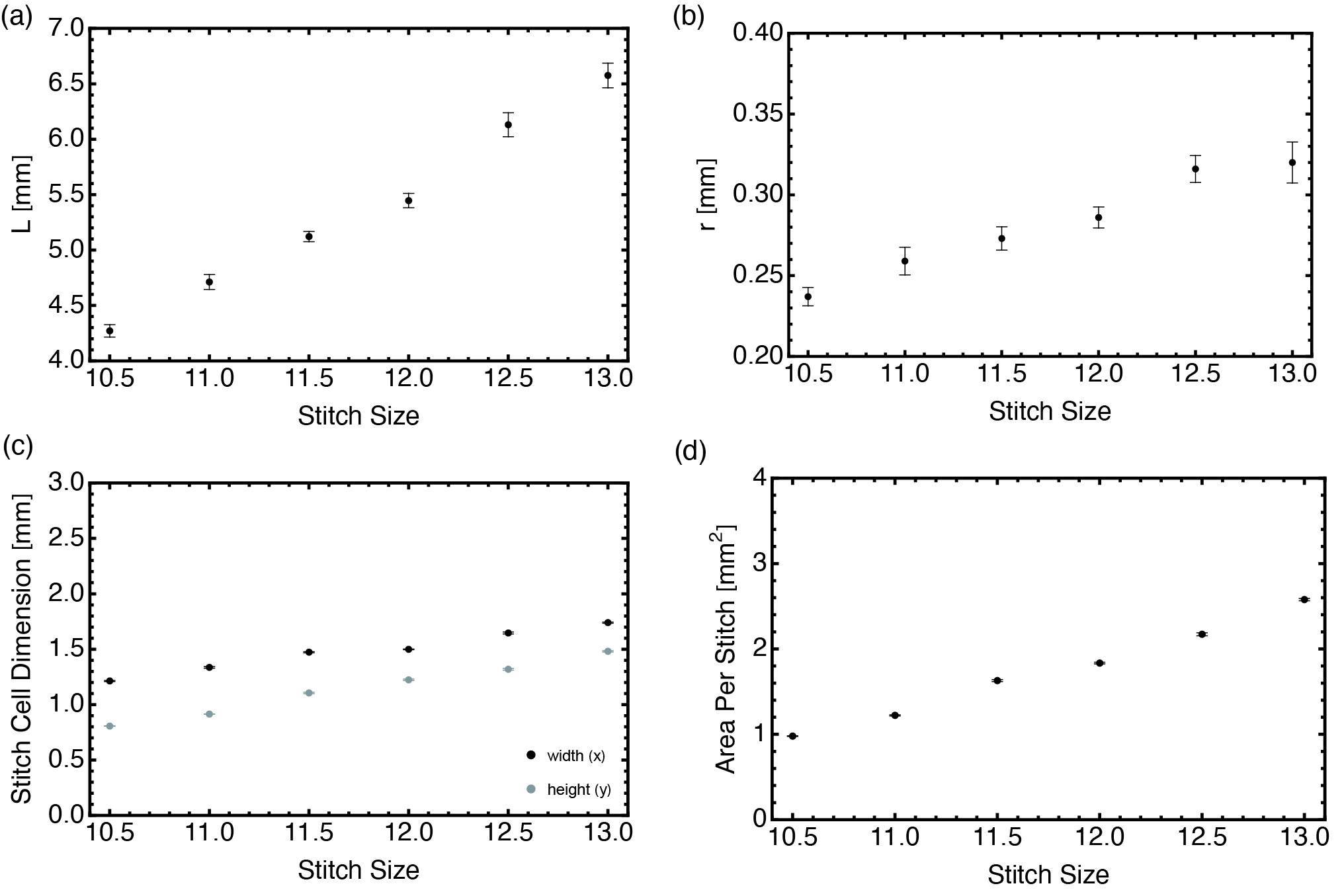}
    \caption{How the (a) length of yarn per stitch, (b) yarn radius, (c) stitch cell dimensions, and (d) area-per-stitch change with the ``stitch size" setting on samples without buffers. Error bars represent the mean standard error.}
    \label{fig:exptension}
\end{figure}

The knitting machine does not have a direct input for the length of yarn per stitch of the fabric.
Instead, the manufacturer indirectly programs in a machine tension via the ``stitch size" parameter.
This tension affects the yarn per stitch and leads to a tighter (smaller ``stitch size") or looser (larger ``stitch size") fabric, shown in Fig. \ref{fig:exptension}a.
The length of yarn per stitch is measured by making fabric samples without the buffer region, with side lengths of 32, 48, and 64 stitches.
The samples are weighed on a scale with 0.01 g precision, and we then use the length to weight ratio of the yarn provided by the yarn manufacturer to calculate an average length of yarn per stitch.
The ``stitch size" also affects the radius of the yarn within the fabric, shown in Fig. \ref{fig:exptension}b.
We measure the yarn radius \textit{in situ} by taking a picture of each fabric sample at rest on the tabletop and measuring the average yarn radius with a digital length and pixel conversion for fifteen different locations within each fabric.
As a consequence of changing length of yarn per stitch and yarn radius, the size of the stitch cell also changes, shown in Fig. \ref{fig:exptension}c,d.
The stitch cell dimensions where taken using a caliper with 0.01 mm precision on samples made without the buffer regions.
Each sample was measured five times and these results are given in \ref{fig:exptension}.

\newpage

\newpage

\clearpage

\section{Simulation Sweeps}

Using simulations allows us to probe the individual effects of a variety of parameters that are difficult to tune with experiment.
Simulation inputs include the yarn radius $r$, the core radius $r_{\rm core}$, the bending modulus $B$, the length of yarn per stitch $L$, the compression energy scaling constant $k$, and the compression energy exponent $p$.

\begin{figure}[h]
    \centering
    \includegraphics[width = 12cm]{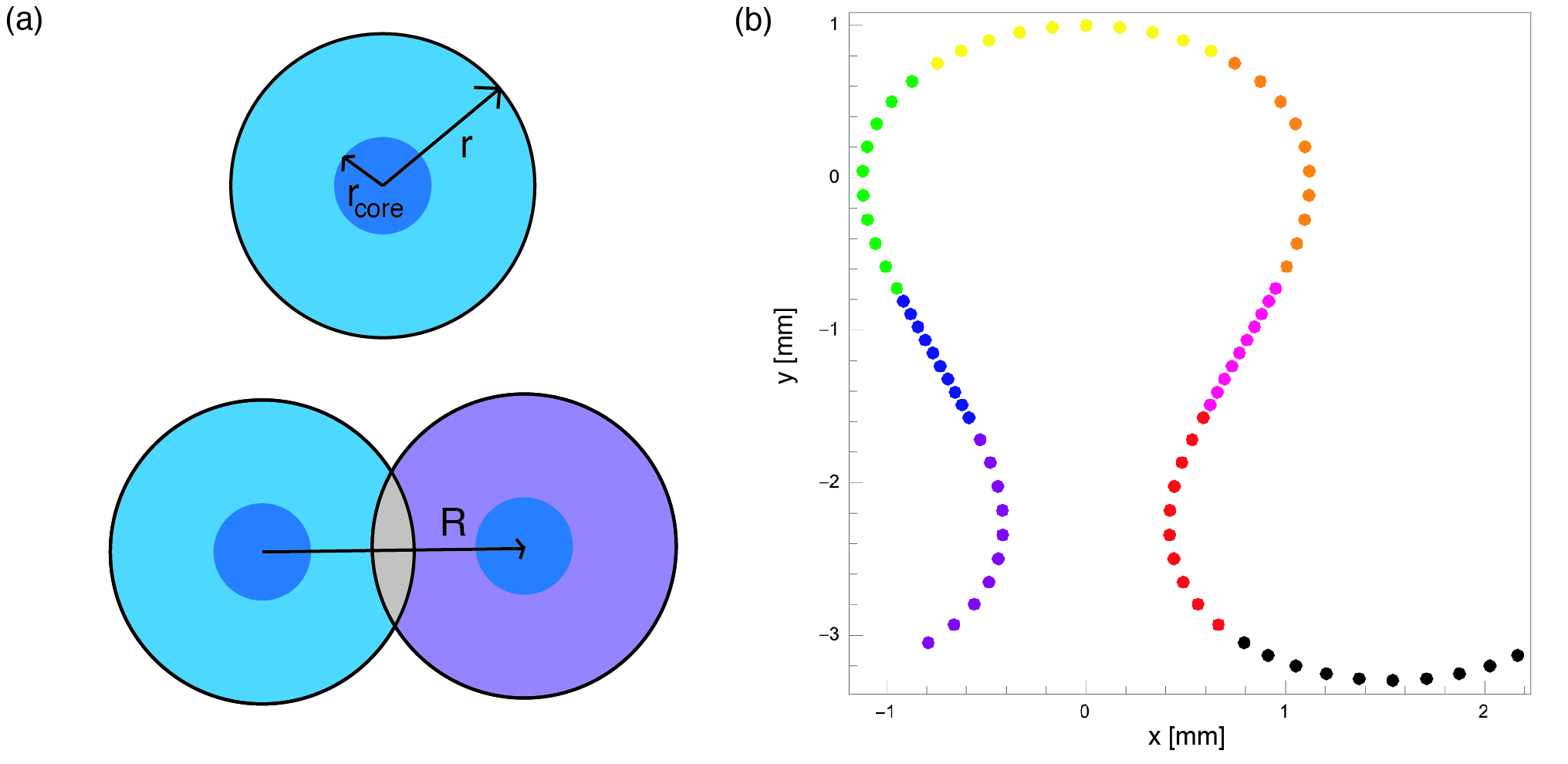}
    \caption{(a) Depiction of yarn using the core-shell model.
    From this view along the length of the yarn, there is a compressible outer region (light blue) of radius $r$ and an incompressible core (dark blue) with radius $r_{\rm core}$.
    The distance between the center-lines of two pieces of yarn is given by $R$.
    In combination, these quantities can describe how far the yarn is compressed (grey region).
    (b) A depiction of the yarn backbone formed by splines.
    Each color segment indicates a different spline.
    For calculating contacts, the splines are sampled with ten points per spline.}
    \label{fig:simfig}
\end{figure}

\begin{figure}[h]
    \centering
    \includegraphics[width = 17.2cm]{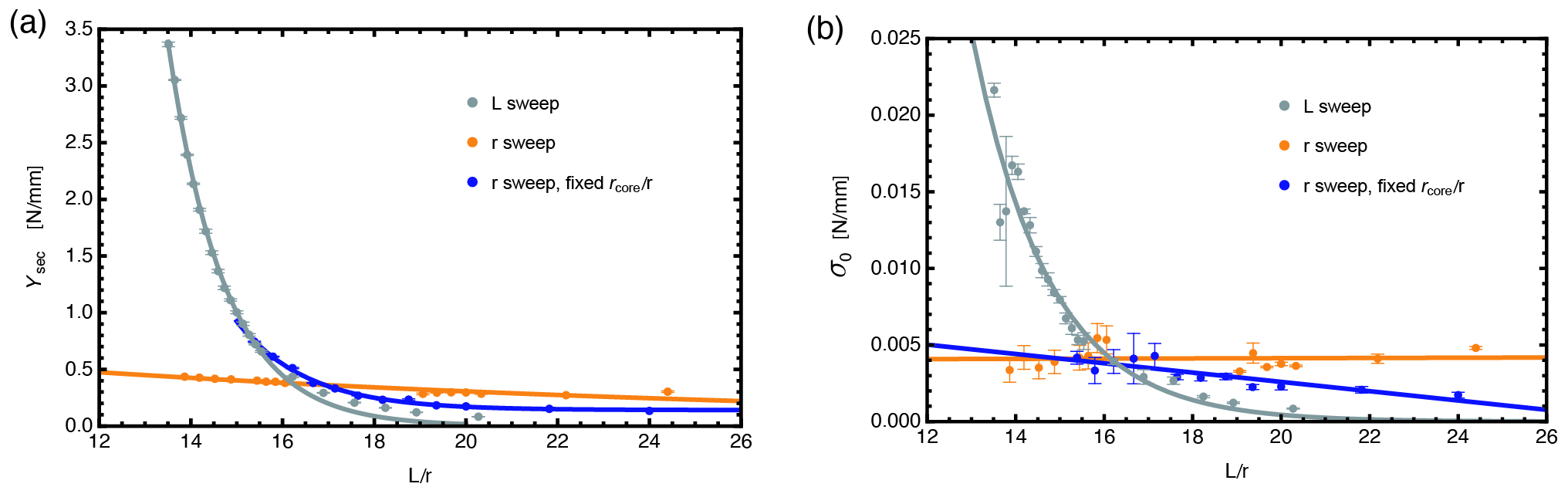}
    \caption{Secant modulus $Y_{\rm sec}$ (a) and offset stress $\sigma_0$ (b) for changing length of yarn per stitch $L$ (gray), yarn radius $r$ (orange), and yarn radius $r$ for fixed ratio $r_{core}/r = 0.45$ (blue) for fabrics stretched in the $y$-direction.
    The length of yarn per stitch is $L=12.2~{\rm mm}$ for the yarn radius sweep and $L=12.0~{\rm mm}$ for the yarn radius with fixed $r_{core}/r$ sweep.
    For the fixed ratio $r_{core}/r$ radius sweep, all values of the yarn radius attempted yielded fabric that is jammed in the $y$-direction.}
    \label{fig:ysimsweeps}
\end{figure}

As discussed in Section \ref{sec:machinetension}, changing the machine tension setting via the ``stitch size" on the industrial knitting machine is confirmed to change the length of yarn per stitch $L$ and the yarn radius $r$.
What happens to the core radius $r_{core}$ is yet unknown and will require further study.
As these parameters change, the jamming metrics $Y_{\rm sec}$ and $\sigma_0$ also change, as shown in Fig. \textcolor{black}{2e,f} in the Main Text for the $x$-direction.
Fig. \ref{fig:ysimsweeps} shows how these metrics change when the fabric is stretched in the $y$-direction.
Fits and fitting parameters are given in \ref{tab:mfitsanderrors}.
All fits were completed using the NonlinearModelFit\cite{NonlinearModelFit} function in Mathematica version 14.0.
Multiple forms were attempted for each curve and the form with the largest $R^2$ value was selected.

\begin{table}[h]
    \centering
    \begin{tabular}{|c|c|c|c|c|}
        \hline
        Stretching Direction & Sweep Parameter & Quantity & Fit & $R^2$ \\
        \hline
        \hline
         x & $L$ & $Y_{\rm sec}$ & $(1.1*10^7\pm2*10^6) e^{(-1.11\pm0.01) L/r}$  & 0.9992 \\
        \hline
        x & $L$ & $\sigma_0$ & $(340000\pm20000) e^{(-1.147\pm0.004) L/r}$  & 0.9999 \\
        \hline
        x & $r$ & $Y_{\rm sec}$ & $(600\pm200)(L/r)^{-2.9\pm0.1}$ & 0.9979 \\
        \hline
        x & $r$ & $\sigma_0$ &  $(0.3\pm0.2)e^{(-0.29\pm0.04)L/r}$  & 0.9712  \\
        \hline
        x & $r$, fixed $r_{core}/r$ & $Y_{\rm sec}$ & $0.01\pm0.06 + (3*10^8\pm3*10^9)(L/r)^{-7\pm3}$  & 0.9525 \\
        \hline
        x & $r$, fixed $r_{core}/r$ & $\sigma_0$ & $(3*10^7\pm6*10^7) e^{(-1.4\pm0.1) L/r}$  & 0.9875 \\
        \hline
        y & $L$ & $Y_{\rm sec}$ & $190000\pm30000 e^{(-0.81\pm0.01) L/r}$  & 0.9992 \\
        \hline
        y & $L$ & $\sigma_0$ & $(47\pm35) e^{(-0.58\pm0.05) L/r}$  & 0.9793 \\
        \hline
         y & $r$ & $Y_{\rm sec}$ & $(0.92\pm0.09) e^{(-0.055\pm0.005) L/r}$ & 0.9965 \\
        \hline
        y & $r$ & $\sigma_0$ &  $0.004\pm0.001 + (1*10^{-6}\pm5*10^{-5})*L/r$  & 0.9761  \\
        \hline
        y & $r$, fixed $r_{core}/r$ & $Y_{\rm sec}$ & $0.141\pm0.009 + (15000\pm8000)e^{(-0.66\pm0.04) L/r}$  & 0.9989 \\
        \hline
        y & $r$, fixed $r_{core}/r$ & $\sigma_0$ & $0.0087\pm0.0010 - (0.00030\pm0.00005)*L/r$  & 0.9832 \\
        \hline
        \hline
        x & Experiment & $Y_{\rm sec}$ & $(3*10^8\pm1*10^9) e^{(-1.1\pm0.2) L/r}$  & 0.9759 \\
        \hline
        x & Experiment & $\sigma_0$ & $(2*10^6\pm1*10^7) e^{(-1.1\pm0.4) L/r}$  & 0.9370 \\
        \hline
        y & Experiment & $Y_{\rm sec}$ & $(4*10^8\pm2*10^9) e^{(-1.0\pm0.2) L/r}$  & 0.9716 \\
        \hline
        y & Experiment & $\sigma_0$ & $(4*10^3\pm2*10^4) e^{(-0.7\pm0.2) L/r}$  & 0.9508 \\
        \hline
    \end{tabular}
    \caption{Fits and errors for $Y_{\rm sec}$ and $\sigma_0$ in both stretching directions for simulation sweeps in the length of yarn per stitch $L$, yarn radius $r$, and yarn radius $r$ for fixed ratio $r_{core}/r$ and the experimental machine tension samples.
    These fits were done using the NonlinearModelFit\cite{NonlinearModelFit} function in Mathematica, version 14.0.
    The fits for the y-direction sweep for yarn radius $r$ with fixed ratio $r_{core}/r$ only represents the behavior of $Y_{\rm sec}$ and $\sigma_0$ within the jammed regime, since none of the sampled parameters resulted in an unjammed fabric.
    All other fits represent some portion of both the jammed and unjammed regimes.}
    \label{tab:mfitsanderrors}
\end{table}

\begin{figure}[h]
    \centering
    \includegraphics[width = 17.2cm]{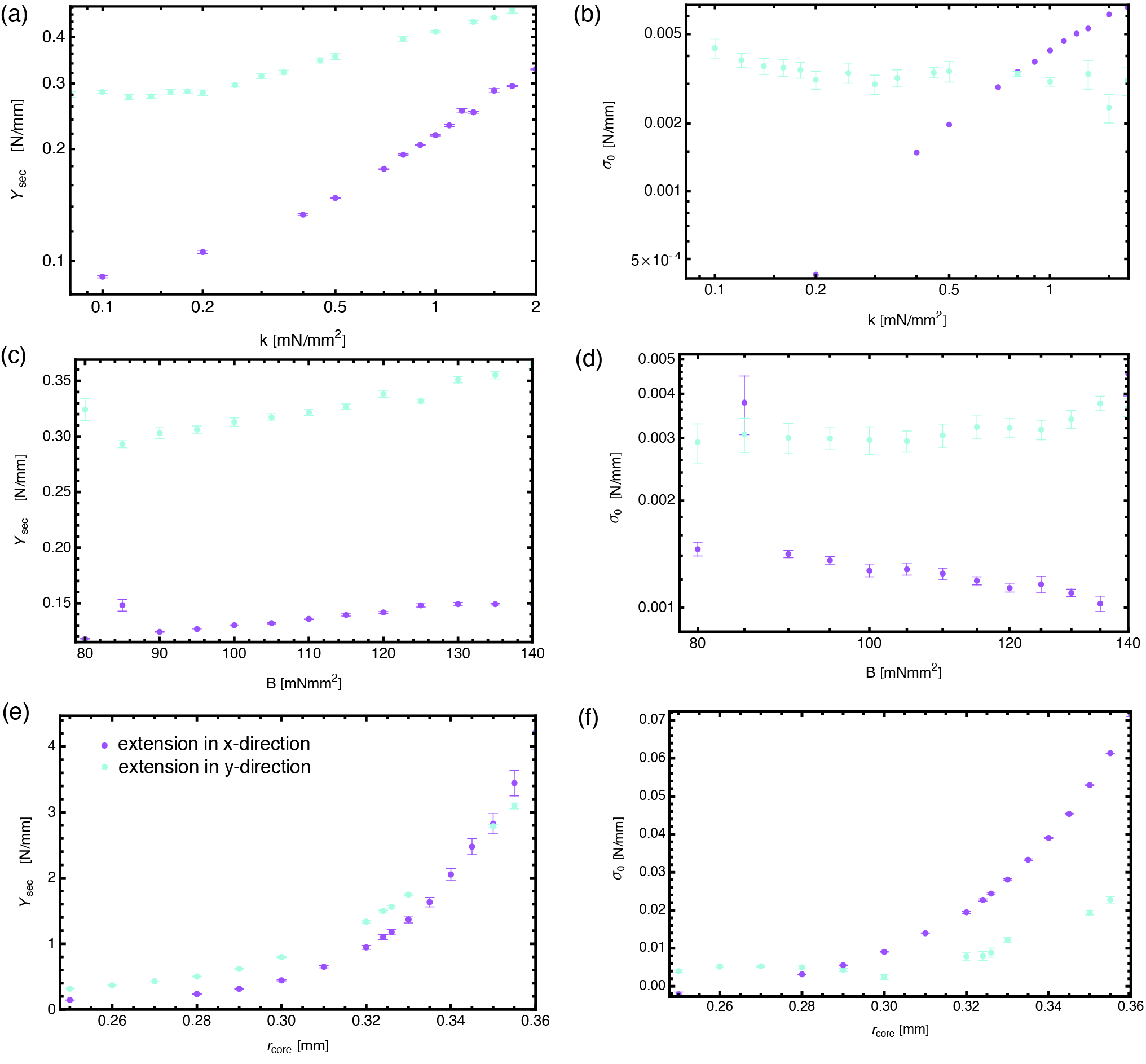}
    \caption{Secant modulus, $Y_{\rm sec}$, of the stress-strain curves for fabrics of different (a) compression scaling constant $k$, (c) bending modulus $B$, and (e) core radii $r_{core}$ pulled in both the $x$- (purple) and $y$-directions (blue).
    Offset stress, $\sigma_0$, of the stress-strain curves for fabrics of different (b) compression scaling constant $k$, (d) bending modulus $B$, and (f) core radii $r_{core}$ pulled in both the $x$- (purple) and $y$-directions (blue).}
    \label{fig:othersweeps}
\end{figure}

In addition to the parameters discussed in the primary paper, there are other simulation parameters that can affect the jammed regime.
The core radius affects the contact potential, described in Main Text Eq. \textcolor{black}{2}.
Figure \ref{fig:othersweeps}e,f shows how the stress-strain jamming metrics, $Y_{\rm sec}$ and $\sigma_0$, change as a function of core radius, which range from $33.7\%$ to $47.3\%$ of the yarn radius (set to 0.74 mm).

\begin{figure}[h]
    \centering
    \includegraphics[width = 17.2cm]{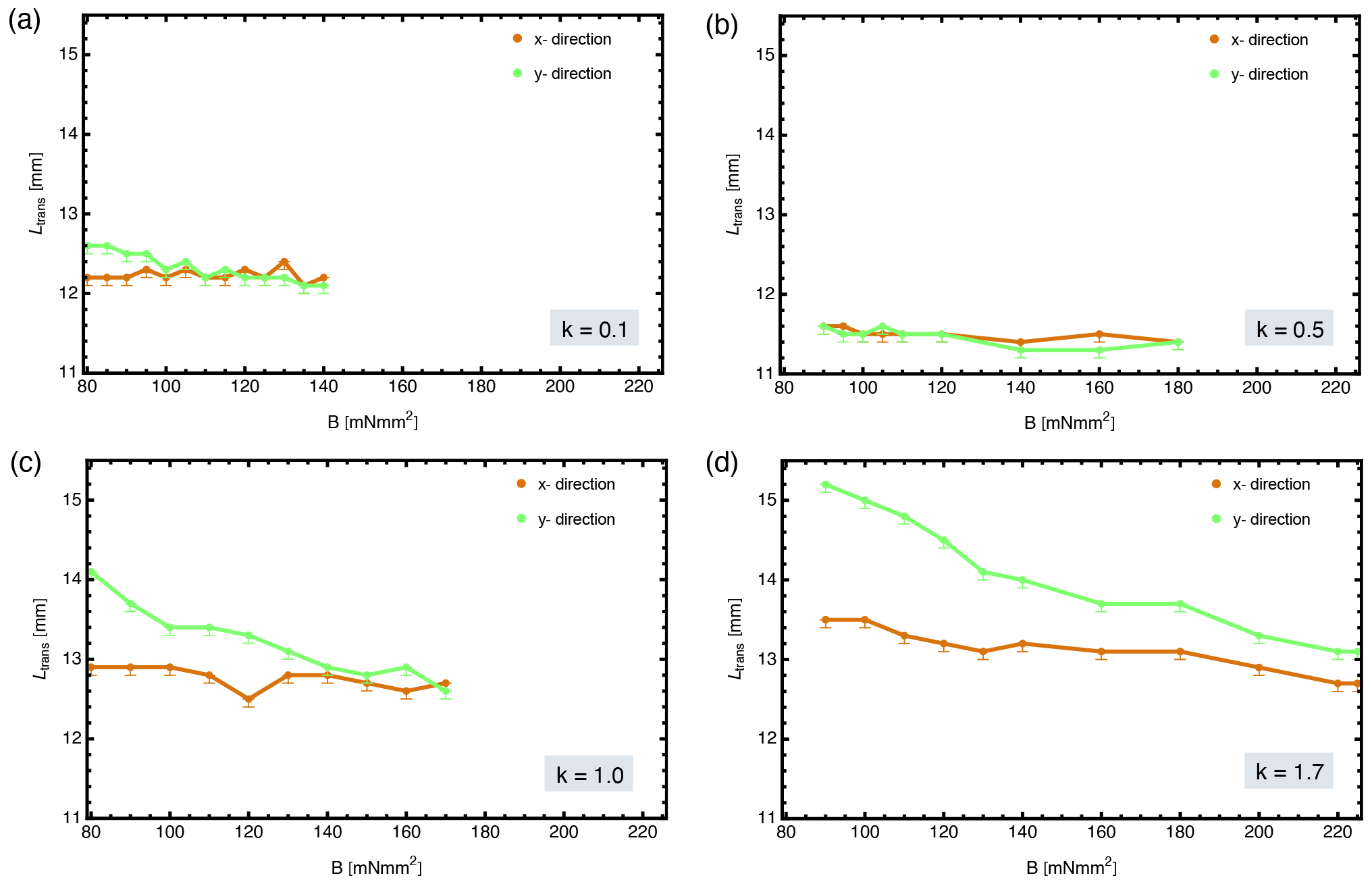}
    \caption{Onset of unjamming $L_{trans}$ as a function of bending modulus for four different values of the compression scaling constant $k$: (a) 0.1 ${\rm mN/mm^2}$, (b) 0.5 ${\rm mN/mm^2}$, (c) 1.0 ${\rm mN/mm^2}$, and (d) 1.7 ${\rm mN/mm^2}$.
    These plots were used to make the composite figure in Main Text Fig. \textcolor{black}{4c}, where $r=0.74~{\rm mm}$.}
    \label{fig:btrans}
\end{figure}

Increasing the bending modulus generally increases jamming in both the $x$- and $y$-directions, shown in \ref{fig:othersweeps}c,d.
Sub-plot (c) shows relatively modest increases in $Y_{\rm sec}$ in comparison to other parameters.
We can also examine the onset of jamming as a function of bending modulus for different values of the compression scaling constant $k$, shown in \ref{fig:btrans}.
Increasing the bending modulus can close the transition gap between the $x$- and $y$-directions, but there is little evidence that is can open a significant gap where the $x$-direction is jammed and the $y$-direction is not.

Figure \ref{fig:othersweeps}a,b shows how the compression scaling constant, $k$, affects the stress-strain jamming metrics.
Extremely small values of $k$ represent the incompressible limit where the yarn is so easy to compress that the core radius acts as an effective yarn radius.
With increasing $k$, the yarn becomes more and more difficult to compress, where infinitely large $k$ would prevent the yarn from compressing at all and represents the second incompressible limit.
As the compression scaling constant increases, the fabric becomes increasingly jammed as shown by increases in the secant modulus in Fig. \ref{fig:othersweeps}a.
The secant modulus shows different power laws for the jammed and unjammed regimes, where jammed fabrics have a secant modulus that increases faster than those for unjammed fabrics. 
The offset stress follows a similar trend for fabrics stretched in the $x$-direction.
For fabrics stretched in the $y$-direction, the fabrics quickly become too stiff to accurately identify a linear regime; the jammed and non-linear regimes become continuous.
This results in a mostly-stable offset stress approximation that does not give good information about the jamming transition.
This is discussed further in Section \ref{sec:byebyelinear}.

\section{Disappearing Linear Elasticity}
\label{sec:byebyelinear}

\begin{figure}[h]
    \centering
    \includegraphics[width = 8.6cm]{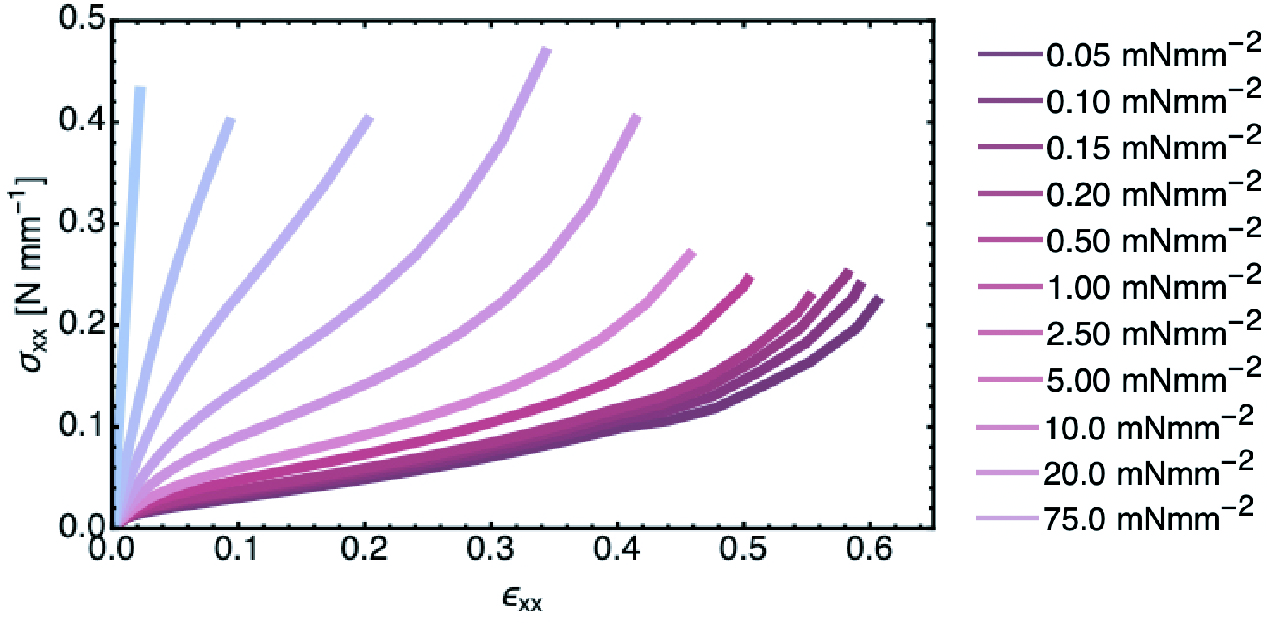}
    \caption{Stress-strain plots for simulations run for different values of the compression scaling constant, $k$, detailed on the right.
    These fabrics were stretched in the $x$-direction and had a length of yarn per stitch of $L=10.7~{\rm mm}$.}
    \label{fig:compstressstrain}
\end{figure}

\begin{figure}[h]
    \centering
    \includegraphics[width = 17.2cm]{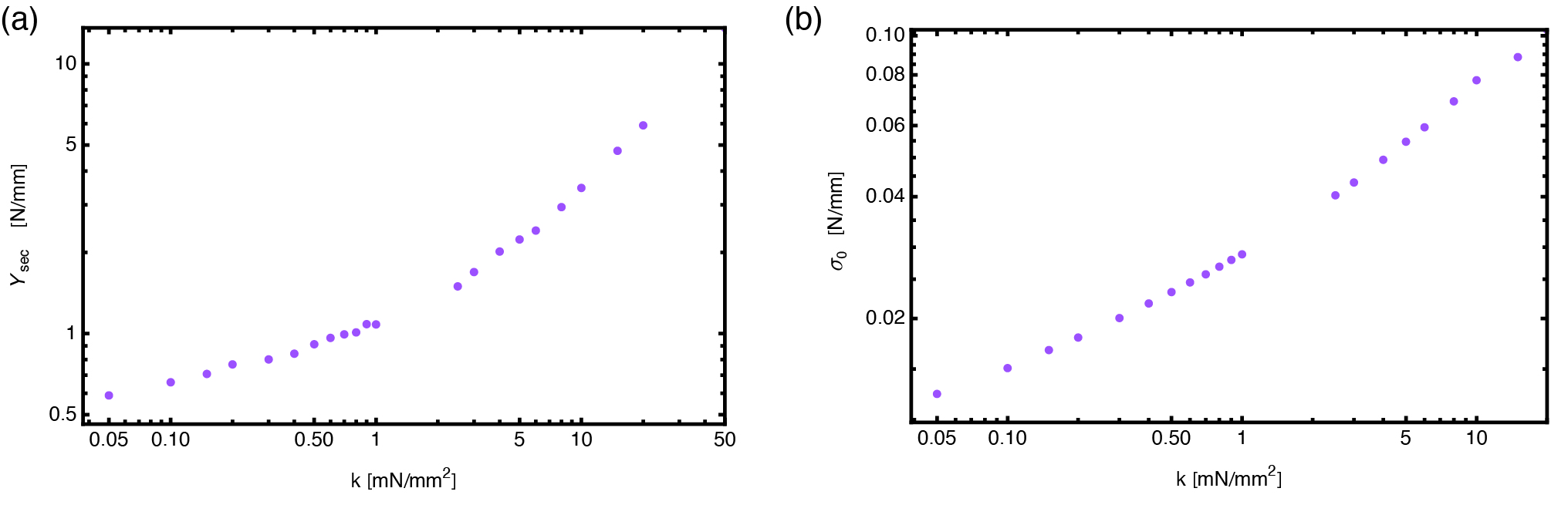}
    \caption{Jamming metric data for the stress-strain curves shown in Fig. \ref{fig:compstressstrain}, all of which are jammed and some of which have little to no perceivable linear elastic regime between jamming and non-linear strain stiffening.
    (a) Secant modulus, $Y_{\rm sec}$, of the stress-strain curves for fabrics of different compression scaling constants pulled in the $x$-direction.
    (b) Offset stress, $\sigma_0$, of the stress-strain curves for fabrics of different compression scaling constants pulled in the $x$-direction.
    The length of yarn per stitch is $L=10.7~{\rm mm}$. Other simulation parameters are as described in Main Text Table \textcolor{black}{1}.}
    \label{fig:compx}
\end{figure}

Fabrics can become so stiff that the linear elastic regime seems to disappear entirely and the fabric moves smoothly from jamming to non-linear strain-stiffening.
This state can be reached by varying any of the simulation parameters sufficiently.
For example, Fig. \ref{fig:compstressstrain} shows how increasing the compression scaling constant $k$ can reduce and then eliminate the linear elastic regime for stiff fabrics.
Fig. \ref{fig:compx} shows how the jamming metrics change with the increasing compression scaling constant.
As seen with the transition from jammed to unjammed, there is a measurable difference in both jamming metrics for these different mechanical regimes.
Moving directly from unjammed fabrics to this very stiff regime in phase space can thus obscure the jamming transition in the jamming metric data.
This is particularly true for the offset stress, $\sigma_0$.
If analyzing data algorithmically with the same bounds for calculating $\sigma_0$, you can observe decreasing $\sigma_0$ as you move into this jammed-to-nonlinear-strain-stiffening regime, as seen in the $y$-extension data for the compression scaling constant sweep data shown in Fig. \ref{fig:othersweeps}b.
This is particularly relevant for extension in the $y$-direction, which is generally stiffer \cite{singal_programming_2024}.

\section{Contact Volumes}

The packing fraction is a useful tool to examine jamming in other systems \cite{unac_effect_2012, Santos2014}.
To determine the compressed volume of the yarn, we use a three-step voxel method.
We first populate the volume surrounding the stitch of interest with voxels of a set side length, $l$.
We determine which voxels are within the volume of the yarn by calculating the distance between the voxel center-point and the stitch center-line.
We eliminate all voxels that are not within one yarn radius of the center-line.
Of this reduced set of voxels $V_{uncomp}$, which represents the un-compressed volume of the yarn within the stitch, we further identify which voxels within this stitch are also contained within a nearby stitch.
This set of voxels represent the compressed regions, $V_{comp}$.
The total volume of the stitch within the fabric is thus

\begin{equation}
    V_{total} = V_{uncomp} - \frac{1}{2}V_{comp}.
\end{equation}

The packing fraction of the stitch is

\begin{equation}
    \phi = \frac{V_{total}}{V_{cell}}
\end{equation}

\noindent
where $V_{cell}$ is the volume of the stitch unit cell.
The stitch unit cell volume is calculated using the cell height and width output of the simulation.
The $z$-dimension is given by $\Delta z = z_{max} - z_{min} + 2*r$ where $z_{max}$ is the maximum $z$-coordinate contained within the stitch center-line, $z_{min}$ is the minimum $z$-coordinate contained within the stitch center-line, and $r$ is the yarn radius.
Voxel data showed that this is an accurate way to calculate the depth of the stitch unit cell that is independent of voxel size.

\begin{figure}[h]
    \centering
    \includegraphics[width = 17.2cm]{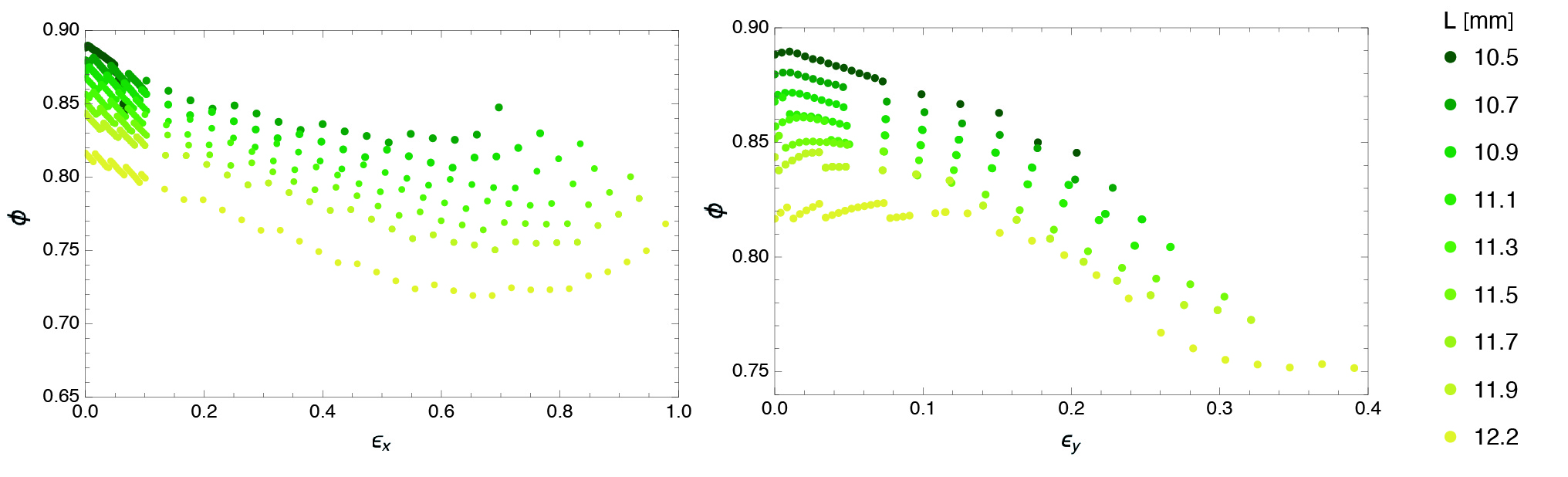}
    \caption{Packing fraction as a function of strain for different lengths of yarn per stitch, $L$.
    Fabrics were extended in both the $x$- (left) and $y$-directions (right).
    Fabrics that display jammed behavior have a length of yarn per stitch $L \leq 11.5~{\rm mm}$.
    For these simulations, the compression scaling constant is $k=0.1~{\rm mN/mm^2}$.}
    \label{fig:pf}
\end{figure}

\begin{figure}[h]
    \centering
    \includegraphics[width = 17.2cm]{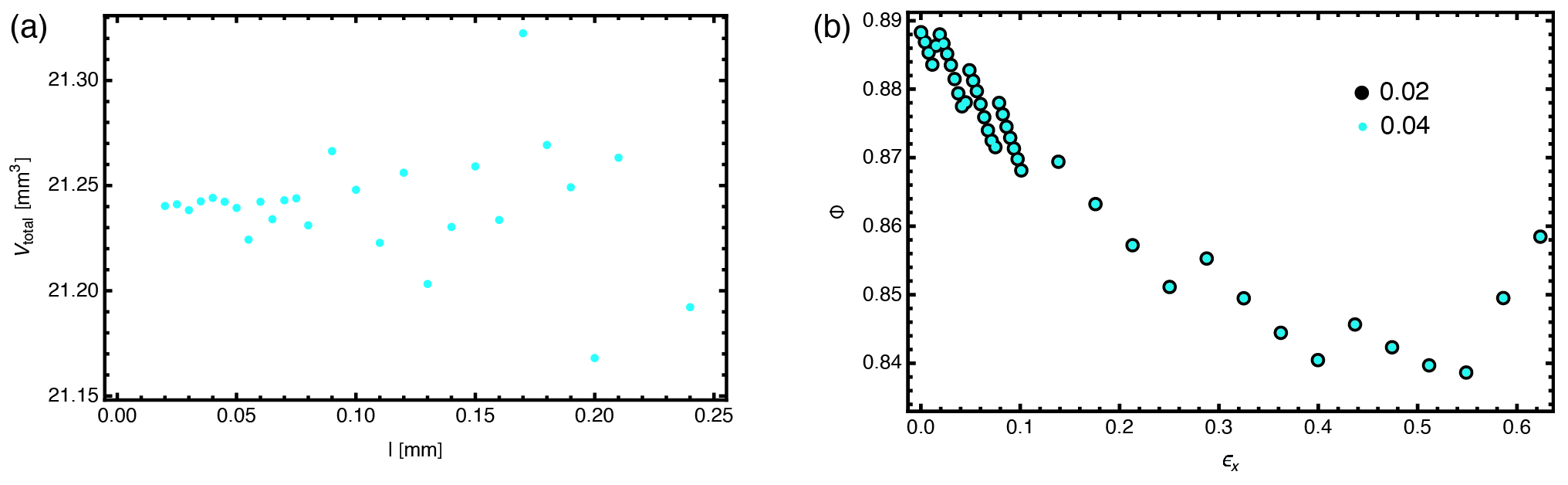}
    \caption{(a) Calculated total volume of the stitch within the fabric using the voxel method as a function of voxel side length.
    Small voxel side lengths ($<0.05$) show stability in the calculated compressed volume of the stitch, whereas larger voxel side lengths show wide variation in volume.
    For these simulations, the length of yarn per stitch is $L=12.2~{\rm mm}$ and the compression scaling constant is $k=0.1~{\rm mN/mm^2}$.
    (b) Packing fraction as a function of strain for fabric stretched in the $x$-direction for two different voxel side lengths: $l=0.04$ in blue and $l=0.02$ in black.
    For these simulations, the length of yarn per stitch is $L=10.5~{\rm mm}$ and the compression scaling constant is $k=0.1~{\rm mN/mm^2}$.}
    \label{fig:voxelsize}
\end{figure}

We calculated the packing fraction as a function of strain for the set of yarn lengths $L \in [10.5, ..., 12.2]$ mm for $k=0.1$ $mN/mm^2$, shown in Fig. \ref{fig:pf}.
From this data, there is no characteristic packing fraction that indicates jamming.
Sweeping over yarn length for zero strain, the packing fraction that indicates the end of jamming is $\phi \approx 0.85$ for this set of yarn parameters, listed in Main Text Table \textcolor{black}{1}.
However, sweeping out in strain for sample $L=10.5~{\rm mm}$, $\phi=0.85$ is at $(\epsilon_x, \epsilon_y)$ location $(0.325,0.177)$, which is far beyond the jamming transition in both extension directions and is even contained within the non-linear strain-stiffening regime.
This indicates that any critical packing fraction in the $L$ phase space is not necessarily critical in strain.
Discontinuities in the packing fraction data make additional analysis, such as transition packing fraction along the strain sweeps as a function of yarn length, unreliable.
These jumps are not the result of the voxel size, shown in \ref{fig:voxelsize}b.
Though there is a maximum voxel side length for reliable volume data (see Fig. \ref{fig:voxelsize}a), significantly reducing this voxel side length did not result in smoother data.
We believe these jumps are the result of small local rearrangements within the stitch.
Since the contact potential is very non-linear and highly dependent on the depth of the contact, the compressed volume of the stitch does not meaningfully represent the energy being minimized in the simulation.
The total volume of the stitch can vary widely without subsequent jumps in the energy of the stitch.
It is possible that a dynamic simulation framework would result in smoother packing fraction data.

\subsection{Contact Forces}

\begin{figure}[h!]
    \centering
    \includegraphics[width = 17.2cm]{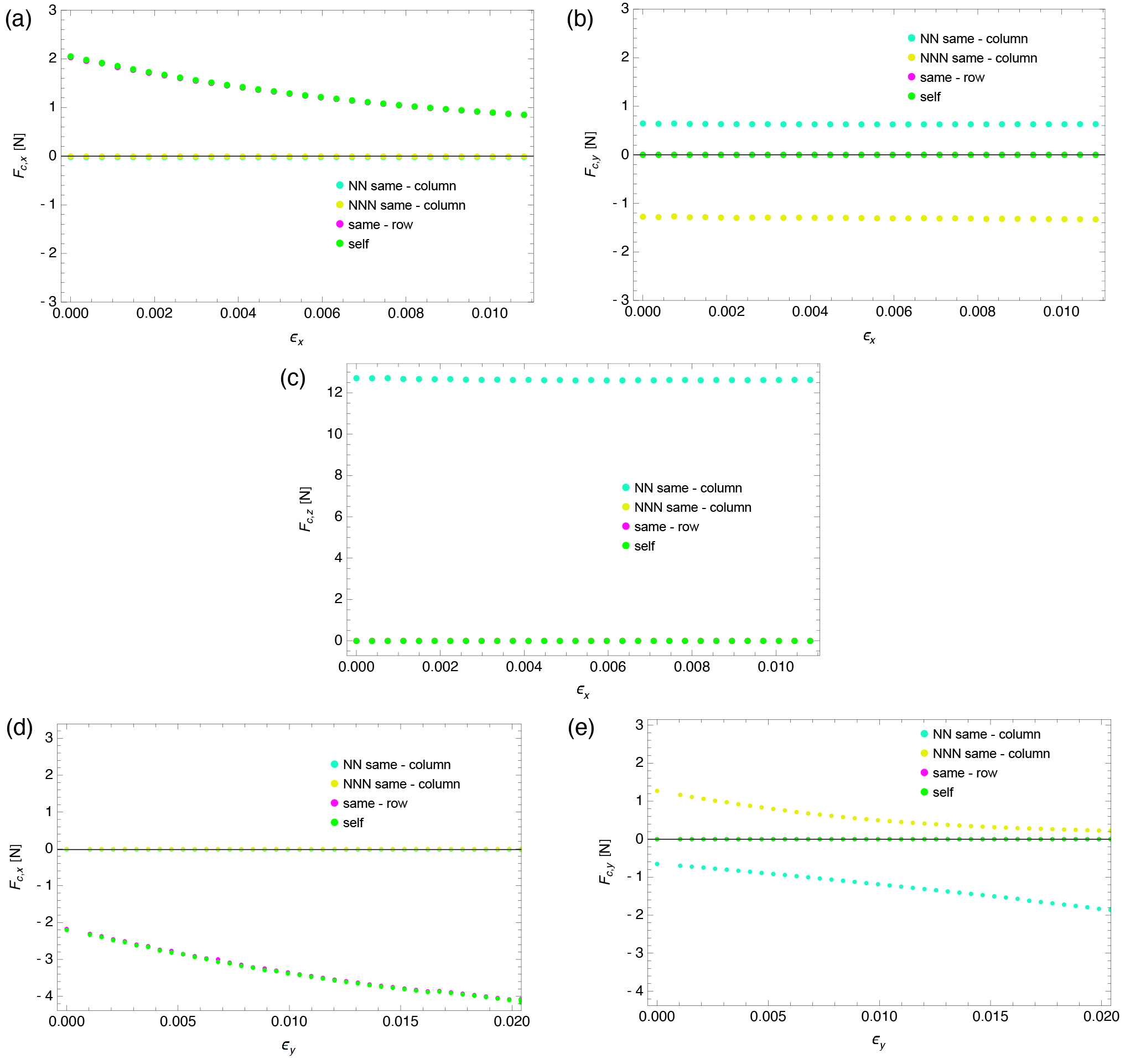}
    \caption{Contact forces as a function of strain in each of the relevant directions for all of the contacts.
    \textbf{The sign of the force indicates whether or not it assists with normal linear elastic behavior (positive) or resists linear elastic mechanics (negative) when stretched in that direction.}
    For each type of contact where relevant, both contacts of that type are added (eg: the NN same-column contact is the sum of this contact force with the stitch above and below).
    Both sets of data are for length of yarn per stitch $L=10.5~{\rm mm}$, which is jammed in both directions.
    (a) Force in the $x$-direction for the fabric stretched in the $x$-direction. The only non-negligible forces are from the same-row (magenta) and self contacts (green), which track together.
    (b) Force in the $y$-direction for the fabric stretched in the $x$-direction.
    The NN same-column contact (blue) promotes typical linear elastic extension while the NNN same-column contact (yellow) resists that motion.
    (c) Force in the $z$-direction (out-of-plane) for the fabric stretched in the $x$-direction.
    The only contribution in this direction is the NN same-column contact (blue), which is pushing out, contrary to the actual motion of the yarn in this jammed configuration.
    (d) Force in the $x$-direction for fabric stretched in the $y$-direction.
    The same-row (magenta) and self (green) contacts are the only contributing contacts, and both are in the direction of expansion.
    (e) Force in the $y$-direction for fabric stretched in the $y$-direction.
    The only contributions are the NN same-column (blue) and NNN same-column (yellow) contacts, which resist linear elastic motion and promote it, respectively.
    The forces in the $z$-direction for fabrics stretched in the $y$-direction are similar to those found when stretching in the $x$-direction, but flipped in sign due to the different out-of-plane linear elastic behavior seen in the $y$-direction (Fig. \ref{fig:zstrain}b).}
    \label{fig:contactforces}
\end{figure}

Each contact volume exerts a force on the yarn within the stitch.
Individual forces for each contact for a single set of yarn and manufacturing parameters are shown in Fig. \ref{fig:contactforces}.
The only non-negligible contact forces in the $x$-direction are the same-row contacts and the self-contact.
Both of these forces are reduced when the fabric extends in the $x$-direction (Fig. \ref{fig:contactforces}a), and thus assist with normal linear elastic behavior when the fabric is pulled in the $x$-direction and inhibit it when the fabric is pulled in the $y$-direction (Fig. \ref{fig:contactforces}d).
The NN same-column contact and the NNN same-column contact are applying forces in the $y$-direction; the NN same-column contact encourages the fabric to shrink in the $y$-direction while the NNN same-column contact encourages the fabric to grow in the $y$-direction.
These forces are stagnant at low strains when the fabric is pulled in the $x$-direction, potentially indicating a lack of contact rearrangement.
When the fabric is pulled in the $y$-direction, the force from the NNN same-column contact decreases in magnitude while the NN same-column contact grows.
The interplay between these two forces and the forces in the $x$-direction determines jamming in the $y$-direction, as discussed in the Main Text and displayed in Fig. \textcolor{black}{6h}.

When the fabric is stretched in the $x$-direction, the mechanics of jamming show the stitch shrinking out-of-plane (Main Text Fig. \textcolor{black}{6g}).
The contact with non-negligible force in this direction is the NN same-column contact, which is applying a force in the direction of expansion, directly contrary to the actual motion of the stitch.
This is additional evidence that bending forces dominate jamming in the $x$-direction. 

\newpage

\newpage

\section{Stitch Cell Dimensions and Area}

\begin{figure}[h]
    \centering
    \includegraphics[width = 17.2cm]{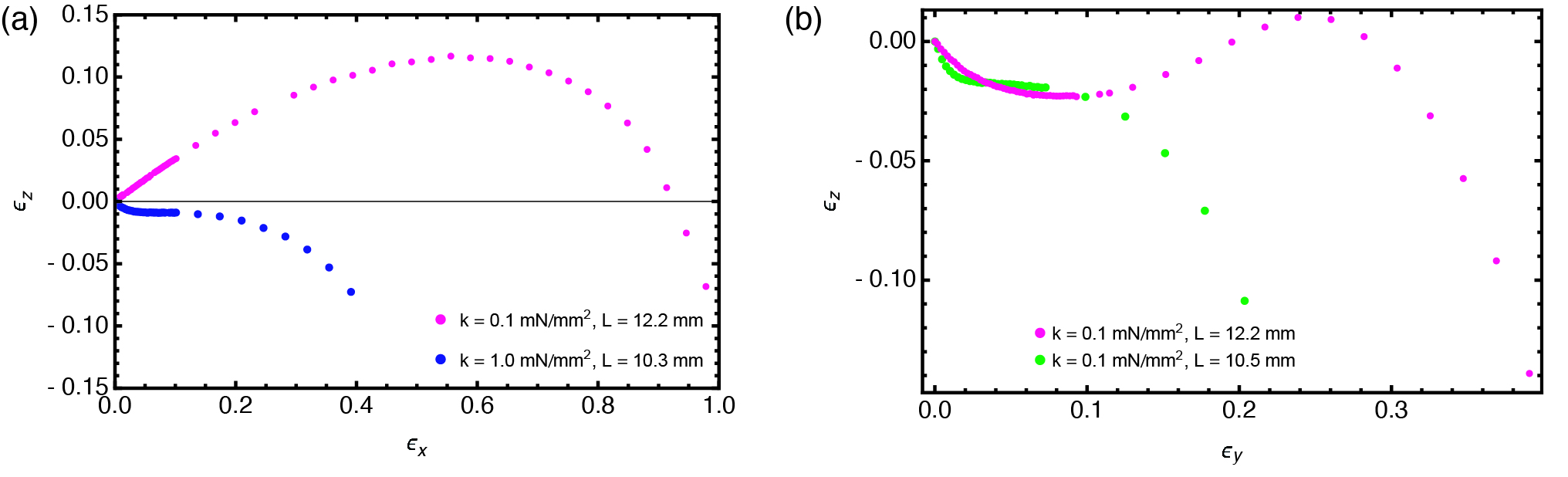}
    \caption{(a) Strain in the $z$-direction as a function of strain in $x$-direction for the full range of motion ($\sigma_x < 0.5$ N/mm) of fabrics stretched in the $x$-direction.
    The fabric in pink is the same unjammed fabric shown in Main Text Fig. \textcolor{black}{6g}.
    The new fabric shown in blue is so jammed that the linear elastic regime has disappeared, described more in Section \ref{sec:byebyelinear}, and maintains a positive Poisson ratio through the entire range of motion.
    (b) Strain in the $z$-direction as a function of strain in $y$-direction for the full range of motion ($\sigma_y < 0.5$ N/mm) of a jammed (green) and unjammed (pink) fabric stretched in the $y$-direction.}
    \label{fig:zstrain}
\end{figure}

As discussed in the Main Text, jammed fabrics pulled in the $x$-direction have unique mechanics markers in the out-of-plane ($z$) strain.
Unjammed fabrics have out-of-plane expansion at low strain and jammed fabrics shrink at low strain.
Fig. \ref{fig:zstrain} shows the full range of motion for one unjammed sample and a superjammed sample.
In the superjammed sample, the out-of-plane width never recovers to the zero-strain dimension through the entirety of the range of motion.
This is due to a lack of a linear elastic regime, which is discussed in more detail in Section \ref{sec:byebyelinear}.

\begin{figure}[h]
    \centering
    \includegraphics[width = 8.6cm]{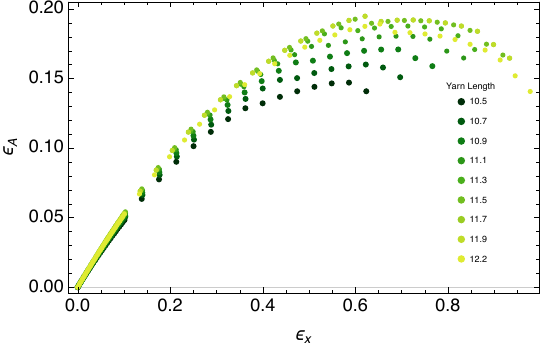}
    \caption{Area strain as a function of strain in the $x$-direction for simulated fabrics of different length of yarn per stitch stretched in the $x$-direction.
    There is no differentiation in area strain between jammed and unjammed fabrics.}
    \label{fig:areastrainx}
\end{figure}

\begin{figure}[h]
    \centering
    \includegraphics[width = 17.2cm]{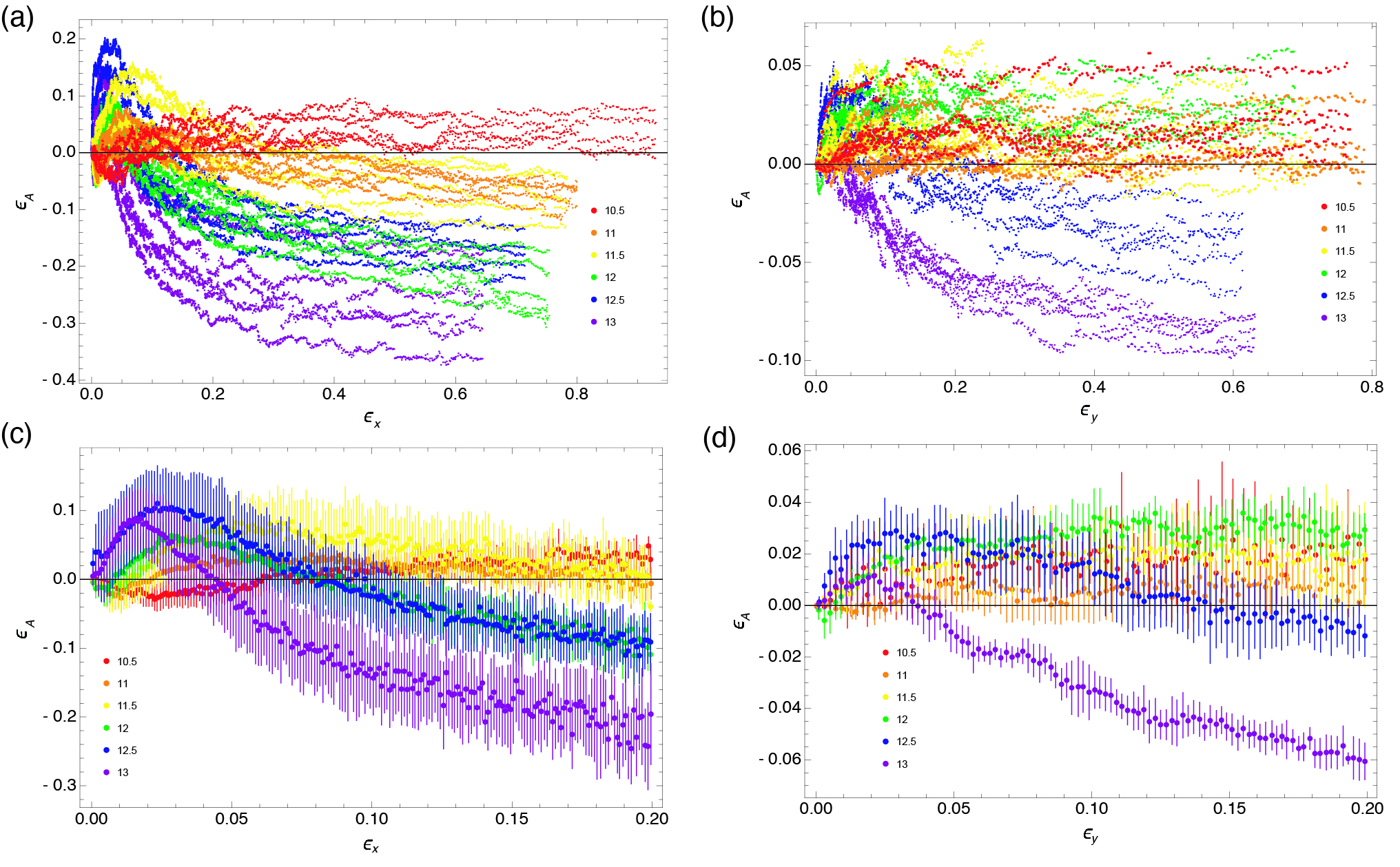}
    \caption{Experimental area strain for different machine tension settings.
    (a) and (b) are raw data for the $x$- and $y$-directions respectively.
    (c) and (d) show average binned area strain data for small strain in the $x$- and $y$-directions respectively, where the bars in each direction are the standard deviations.
    One bin spans $\epsilon_x=0.001$ and $\epsilon_y=0.002$ and these sizes were determined based on the number of data points in the raw experimental data.
    Each fabric has a different number of data points in each strain range since the experimental procedure fixes the extension rate and the fabrics have different elasticities and rest lengths.}
    \label{fig:expareastrain}
\end{figure}

For simulated fabrics pulled in the $x$-direction, the area-per-stitch strain is non-monotonic, first increasing and then beginning to decrease at very large strain as seen in \ref{fig:areastrainx}.
Unlike fabrics stretched in the $y$-direction, there is no marker of jamming in the simulated area strain for fabrics stretched in the $x$-direction.
Though we cannot directly compare these results to experimental area strain, seen in \ref{fig:expareastrain}, because of the ``stitch size" affect on yarn radius and potentially the core radius, both the simulations and experiments show periods of increasing area strain followed by decreasing area strain.
The simulations also do not consider boundary effects, which would provide large constraints on the cell area, particularly at high strain.
These boundary effects may be the cause of the relatively constant area strain at large uniaxial strain, which is not present in simulations.

\section{Tolerance and Error}

In each simulation, we hard-code the energy tolerance to $10^{-5}$ J.
In our specific minimization method, Sequential Least Squares Programming, this tolerance acts as the precision goal for the total energy functional that is being minimized \cite{SciPymin}.
For a single set of simulation parameters, including a determined set of stitch cell dimensions, we establish the error in energy as $10^{-5}$ J according to this simulation tolerance.
To determine affects of our discrete set of stitch cell dimensions on measured quantities, such as the energy gap, we chose an un-jammed simulation set for both the $x$- and $y$-directions, did a linear fit to a small subset of the contact energy data points, and computed the root mean square.
Results are summarized in \ref{tab:gaperror}.
Linear fits were conducted with the Fit\cite{Fit} function in Mathematica version 14.0.
We estimate the error in all energy gap analysis as $5*10^{-5}$ J.

\begin{table}[h]
    \centering
    \begin{tabular}{|c|c|c|c|c|c|}
        \hline
        extension direction & $k$ [mN/mm\textsuperscript{2}] & $L$ [mm] & Step Size [mm] & $\epsilon_{final}$ & RMS [J] \\
        \hline
        \hline
        x & 1 & 14.0 & 0.01 & 0.056 & $2.27*10^{-5}$ \\
        x & 1.5 & 14.0 & 0.01 & 0.075 & $4.71*10^{-5}$ \\
        x & 2 & 15.0 & 0.01 & 0.052 & $3.68*10^{-5}$ \\
        y & 1 & 14.4 & 0.002 & 0.029 & $2.51*10^{-5}$ \\
        y & 1.5 & 15.4 & 0.01 & 0.034 & $3.92*10^{-5}$ \\
        y & 2 & 16.3 & 0.002 & 0.028 & $3.06*10^{-5}$ \\
        \hline
    \end{tabular}
    \caption{Results for energy gap error analysis.
    Step size indicates how much the stitch cell size increased in the extension direction in the low-strain regime during the simulation.
    $\epsilon_{final}$ is the largest strain value contained in the subset of data points used for this error analysis.}
    \label{tab:gaperror}
\end{table}

For the contact force, we repeat this analysis on the $z$-component of the same-row contact for a variety of yarn lengths with a scaling compression constant of $k=0.1~{\rm mN/mm^2}$.
This contact force is relatively linear and consistent for both jammed and unjammed fabrics.
These simulations were stretched in the $y$-direction and all have the same step size, $\Delta a_y = 0.001~{\rm mm}$, and each RMS was calculated for the first 20 force vs strain points, displayed in \ref{tab:forceerror}.
The average RMS over all yarn lengths is $2.24*10^{-5}$ N.

\begin{table}[h]
    \centering
    \begin{tabular}{|c|c|c|c|c|}
        \hline
        $L$ [mm] & $L/r$ & $|F_{x}|$ [N] & $|F_{y}|$ [N] & RMS [N]  \\
        \hline
        \hline
        10.9 & 14.73 & 1.952 & 0.565 & $6.89*10^{-5}$ \\
        11.0 &  14.86 & 1.588 & 0.569 & $2.30*10^{-5}$ \\
        11.1 &  15.0 & 1.199 & 0.548 & $1.44*10^{-5}$ \\
        11.2 & 15.14 & 0.956 & 0.523 & $3.04*10^{-5}$ \\
        11.3 & 15.27 & 0.737 & 0.520 & $1.45*10^{-5}$ \\
        11.4 & 15.41 & 0.558 & 0.487 & $3.63*10^{-5}$ \\
        11.5 & 15.54 & 0.423 & 0.462 & $2.72*10^{-5}$ \\
        11.6 & 15.68 & 0.321 & 0.451 & $2.41*10^{-5}$ \\
        11.7 & 15.81 & 0.245 & 0.430 & $2.01*10^{-5}$ \\
        11.8 & 15.91 & 0.186 & 0.410 & $1.44*10^{-5}$ \\
        11.9 & 16.08 & 0.152 & 0.384 & $8.30*10^{-6}$ \\
        12.0 & 16.22 & 0.118 & 0.380 & $5.69*10^{-5}$ \\
        12.8 & 17.30 & 0.028 & 0.405 & $2.93*10^{-6}$ \\
        \hline
    \end{tabular}
    \caption{Results for contact force error analysis.
    The RMS was calculated on the $z$-component of the force for the same row contact.
    Generally, the $z$-component of the contact force was the most linear component.
    Data in this table was used to make Main Text Fig. \textcolor{black}{6h}.}
    \label{tab:forceerror}
\end{table}

Linear fits to the jammed regime of the energy gap data were used in computing $\Delta L/r$, the length regime where the fabric transitions from jammed to unjammed configurations.
The linear fits for each set of energy gap data along with fitting errors are described in \ref{tab:linfiterrors}.
Fits were conducted with the Fit\cite{Fit} function in Mathematica version 14.0.
The mean square error for a majority of the linear fits (12 of 17) falls within the error of the energy gap calculation.

\begin{table}[h]
    \centering
    \begin{tabular}{|c|c|c|c|c|c|}
        \hline
        $k$ [mN/mm\textsuperscript{2}] & Slope [J/mm] & Y-Intercept [J] & Max Abs(Residual) [J] & Mean Abs(Residual) [J] & MSE [J\textsuperscript{2}]  \\
        \hline
        \hline
        0.001 & $-6.99*10^{-4}$ & 0.0077 & $6.24*10^{-5}$ & $2.92*10^{-5}$ & $1.25*10^{-9}$\\
        0.1 & $-1.55*10^{-3}$ & 0.0176 & $1.71*10^{-5}$ & $6.54*10^{-6}$ & $6.97*10^{-11}$\\
        0.2 & $-2.00*10^{-3}$ & 0.0230 & $2.94*10^{-5}$ & $1.47*10^{-5}$ & $3.12*10^{-10}$\\
        0.4 & $-2.64*10^{-3}$ & 0.0306 & $5.54*10^{-5}$ & $2.08*10^{-5}$ & $7.35*10^{-10}$\\
        0.5 & $-2.88*10^{-3}$ & 0.0334 & $3.74*10^{-5}$ & $1.19*10^{-5}$ & $3.09*10^{-10}$\\
        0.7 & $-2.91*10^{-3}$ & 0.0343 & $5.32*10^{-5}$ & $2.99*10^{-5}$ & $1.14*10^{-9}$\\
        0.8 & $-3.00*10^{-3}$ & 0.0355 & $8.37*10^{-5}$ & $4.61*10^{-5}$ & $2.69*10^{-9}$\\
        0.9 & $-3.12*10^{-3}$ & 0.0370 & $7.05*10^{-5}$ & $3.63*10^{-5}$ & $1.75*10^{-9}$\\
        1.0 & $-3.03*10^{-3}$ & 0.0361 & $1.08*10^{-4}$ & $4.68*10^{-5}$ & $3.34*10^{-9}$\\
        1.1 & $-3.15*10^{-3}$ & 0.0376 & $8.85*10^{-5}$ & $5.19*10^{-5}$ & $3.30*10^{-9}$\\
        1.2 & $-3.27*10^{-3}$ & 0.0390 & $6.64*10^{-5}$ & $2.62*10^{-5}$ & $9.87*10^{-10}$\\
        1.3 & $-3.10*10^{-3}$ & 0.0373 & $1.02*10^{-4}$ & $3.86*10^{-5}$ & $2.20*10^{-9}$\\
        1.5 & $-3.13*10^{-3}$ & 0.0380 & $9.79*10^{-5}$ & $4.65*10^{-5}$ & $2.99*10^{-9}$\\
        1.7 & $-2.83*10^{-3}$ & 0.0349 & $7.26*10^{-5}$ & $3.97*10^{-5}$ & $2.30*10^{-9}$\\
        2.0 & $-2.91*10^{-3}$ & 0.0361 & $1.29*10^{-4}$ & $4.94*10^{-5}$ & $4.54*10^{-9}$\\
        3.0 & $-2.92*10^{-3}$ & 0.0371 & $5.48*10^{-5}$ & $2.71*10^{-5}$ & $1.00*10^{-9}$\\
        4.0 & $-2.49*10^{-3}$ & 0.0328 & $9.39*10^{-5}$ & $3.14*10^{-5}$ & $1.91*10^{-9}$\\
        \hline
    \end{tabular}
    \caption{Fitting information for the linear fits to the jammed regime of the contact energy gap data for various values of the compression scaling constant $k$.
    The maximum value of the absolute value of the residuals and the mean of the absolute value of the residuals are listed.
    The mean absolute value of the residuals is within our estimated energy gap error for all but one linear fit ($k=1.1 {\rm mN/mm^2}$).
    The mean squared error is within our estimated energy gap error for all but five values of $k$ ($k = {0.8,1.0,1.1,1.5,2.0}$).}
    \label{tab:linfiterrors}
\end{table}

\begin{figure}[h]
    \centering
    \includegraphics[width = 17.2cm]{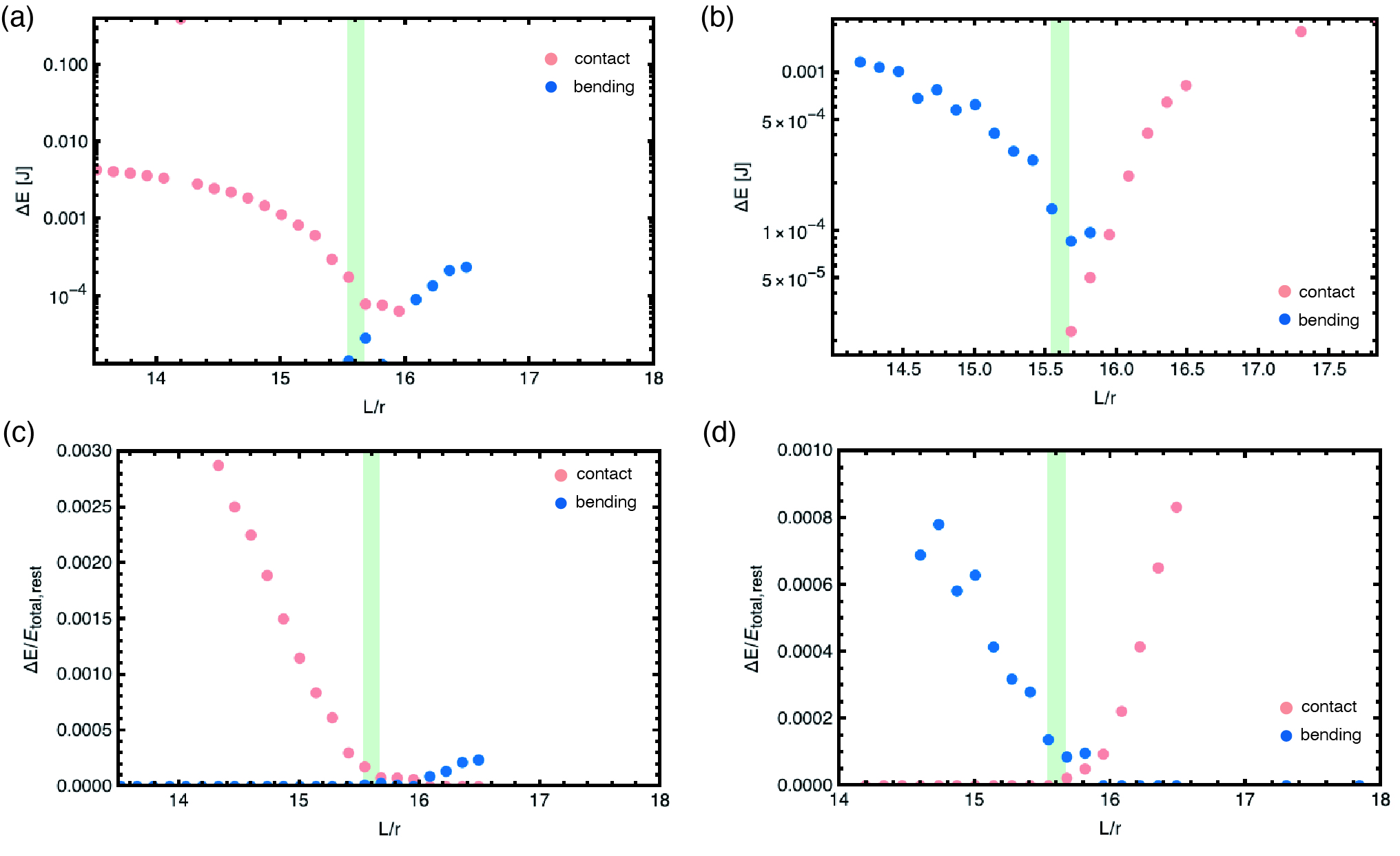}
    \caption{Energy gap for the contact (coral) and bending (blue) energies as a function of length of yarn per stitch for compression scaling constant $k=0.1$ mN/mm\textsuperscript{2} pulled in the $x$-direction (a) and the $y$-direction (b) on a log scale.
    This is a re-scaling of Main Text Fig. \textcolor{black}{3}.
    All simulations where the energy gap is zero are not present on this graph due to the logarithmic scaling.
    The light green region shows where in length of yarn per stitch parameter space the fabric moves from the jammed transition region ($\Delta E_c > 5*10^{-5}$ J) to un-jammed ($\Delta E_c < 5*10^{-5}$ J).
    We can see a change in behavior in the jammed energy gap when $L/r$ approaches the green end-of-transition region, supporting our error floor of $5*10^{-5}$ J.
    (c,d) Energy gap rescaled by the total energy of the rest, zero-force configuration in the (c) $x$- and (d) $y$-directions.
    This provides an intrinsic view of the extrinsic definition of the energy gap (Main Text Eq. \textcolor{black}{4}).
    }
    \label{fig:LerrorLog}
\end{figure}

We define the end of the jamming transition region when the energy gap is less than the energy gap error defined above, $5*10^{-5}$ J.
The true length of yarn per stitch for the end of the transition region would lie at this energy gap error boundary.
Due to the size of our length sweep, 0.1 mm, all of the transition lengths reported are beneath this energy gap error boundary.
The true end of the transition region is thus between the reported transition length and the next largest length of yarn per stitch, shown in the green regions of Main Text Fig. \textcolor{black}{3} and Fig. \ref{fig:LerrorLog}.
The error for each of our reported transition lengths is thus asymmetric and plus zero, minus 0.1 mm ($\pm _{0.1}^0$ mm).

\section{Criticality and Fabrics that Never Jam}

The phase space diagram in Fig. \textcolor{black}{1c} indicates the presence of a critical point, $(L/r)^*$.
Fabrics with $L/r > (L/r)^*$ do not show jammed behavior for any applied stress.
$(L/r)^*$ thus separates fabrics into two regimes, fabrics that can jam and fabrics that can't.
Our simulations support $(L/r)^*$ as a finite point, but do not allow us to determine it's value.
The energy gap data in Fig. \ref{fig:LerrorLog} shows that the energy gap doesn't just approach zero asymptotically, but converges to zero.
The change in behavior for energy gap values above and below the energy gap error floor ($5*10^{-5}$ J) indicates that there may be a scaling law that describes this critical point $(L/r)^*$, but this data is insufficient to accurately describe it.

\end{document}